\documentclass[11pt]{article}
\usepackage[letterpaper,margin=1in]{geometry}
\usepackage[all,pdf]{xy}
\usepackage{amsfonts}
\usepackage{lscape}
\usepackage{amssymb}
\usepackage{bbm}
\usepackage{color}
\usepackage{amsfonts,amssymb,mathrsfs,amsmath,amssymb,theorem,float}
\usepackage{graphicx}  

\newtheorem{theorem}{Theorem}[section]

\newtheorem{cor}[theorem]{Corollary}
\newtheorem{lemma}[theorem]{Lemma}
\newtheorem{alg}[theorem]{Algorithm}
\newtheorem{remark}[theorem]{Remark}
\newtheorem{defi}[theorem]{Definition}

\def\qed{\hfil {\vrule height5pt width2pt depth2pt}}

\def\qed{\hfil {\vrule height5pt width2pt depth2pt}}
\def\bref#1{(\ref{#1})}

\def\qed{\hfil {\vrule height5pt width2pt depth2pt}}
\def\S{\mathcal{S}}

\def\proof{{\noindent\em Proof.\,\,}}

\def\bref#1{(\ref{#1})}

\def\N{{\mathbb N}}
\def\Z{{\mathbb Z}}

\def\X{{\mathbb{X}}}

\def\PP{{\mathcal P}}

\def\F{{\mathbb {F}}}

\def\RB{{\mathcal{R}}}

\def\bref#1{(\ref{#1})}

\def\+{ \oplus}
\def\-{\ominus}
\def\*{\otimes}

\def\deg{\hbox{\rm{deg}}}

\def\modp{{\mathbf{mod}}}

\begin{document}

\title{Faster Interpolation Algorithms for Sparse Multivariate Polynomials Given by Straight-Line Programs\thanks{Partially
       supported by a NSFC grant No.11688101.}}
\author{Qiao-Long Huang$^{1,2}$ and Xiao-Shan Gao$^{1,2}$ \\
 $^{1}$KLMM, Academy of Mathematics and Systems Science,
 Chinese Academy of Sciences\\
 $^{2}$University of Chinese Academy of Sciences}
\date{}

\maketitle

\begin{abstract}
\noindent
In this paper, we propose new deterministic and Monte Carlo interpolation algorithms for sparse multivariate polynomials represented by straight-line programs.
Let $f$ be an $n$-variate polynomial given by a straight-line program, which has a degree bound $D$ and a term bound $T$. Our deterministic algorithm is quadratic in $n,T$ and cubic in $\log D$ in the Soft-Oh sense,
which has better complexities than existing deterministic interpolation algorithms in most cases.
Our Monte Carlo interpolation algorithms have better complexities than existing Monte Carlo interpolation algorithms and are the first algorithms whose complexities are linear in $nT$  in the Soft-Oh sense. Since $nT$ is a factor of the size of $f$, our Monte Carlo algorithms are optimal in $n$ and $T$ in the Soft-Oh sense.
\end{abstract}

\section{Introduction}

The sparse interpolation for multivariate polynomials has received considerable interest.
There are two basic models for this problem:
the polynomial is either given as a straight-line program (SLP) \cite{5c,5,8,5d}
or a more general black-box \cite{5a,3,6,813,13}.
%
In this paper, we consider the problem of  interpolation for a sparse multivariate polynomial given by an SLP.

\subsection{Main results}

Let $f\in\RB[x_1,\ldots,x_n]$ be an SLP polynomial of length $L$, with a degree bound $D$ and a term bound $T$, where $\RB$ is a computable ring.
%
In this paper, all complexity analysis relies on the ``Soft-Oh" notation  $O^\thicksim(\phi)=O(\phi\cdot \mathbf{polylog}(\phi))$, where $\mathbf{polylog}$ means $\log^c$ for some fixed $c>0$. When we say ``linear", ``optimal'', etc., we mean linear and optimal in the sense of ``Soft-Oh" complexity.

We propose a new deterministic interpolation algorithm for $f$ given by an SLP, whose complexity is $O^\thicksim(Ln^2T^2\log^2 D+LnT\log^3 D)$ $\RB$ arithmetic operations and a similar number of extra bit operations.
We also propose two new Monte Carlo interpolation algorithms for SLP multivariate polynomials.
For a given $\mu\in(0,1)$, the complexity of our first algorithm is $O^\thicksim(LnT(\log^3 D+\log D\log\frac1\mu))$ $\RB$ arithmetic operations, and with probability at least $1-\mu$, it returns the correct polynomial.
In our second algorithm, $\RB$ is a finite field $\F_q$ and we can evaluate $f$ in a proper extension field of $\F_q$. The bit complexity of our second algorithm is $O^\thicksim(LnT\log^2 D(\log D+\log q)\log\frac1\mu)$, and with probability at least $1-\mu$, it returns the correct polynomial.
These algorithms are the first ones whose complexity is linear and optimal in $n$ and $T$.

\begin{table}[ht]
\footnotesize
\centering
\begin{tabular}{c|c|c}
Algorithms &Total Cost & Type\\ \cline{1-3}
Dense& $LD^n$&Deterministic\\
%
Garg $\&$ Schost \cite{5} &$Ln^2T^4\log^2D$& Deterministic\\
Randomized G $\&$ S \cite{8}&$Ln^2T^3\log^2D$& Las Vegas\\
Arnold, Giesbrecht $\&$ Roche \cite{2} &$Ln^3T\log^3 D$& Monte Carlo\\ \cline{1-3}
This paper (Thm \ref{the-2}) &$Ln^2T^2\log^2 D+LnT\log^3 D$& Deterministic\\
This paper (Thm \ref{them-5})&$LnT\log^3 D$&Monte Carlo
\end{tabular}
\caption{A ``soft-Oh" comparison  for SLP  polynomials over an arbitrary ring $\RB$}\label{tab-1}
\end{table}

In Table \ref{tab-1}, we list the complexities for the SLP interpolation algorithms which are similar to the methods proposed in this paper.
In the table, $L$ is the size of the SLP  and ``total cost" means the number of arithmetic operations in $\RB$.
For deterministic algorithms,
the ratio of the cost of our deterministic method
with that of the algorithm given in \cite{5} is $\frac{nT+\log D}{nT^3}$.
%
So our method has better complexities unless  $f$ is super sparse,
more precisely, unless $T < \sqrt[3]{\frac{\log D}{n}}
$ in the Soft-Oh sense.
Noting that a dense polynomial with degree $d$ has $\left( \begin{array}{c}d\cr d+n \end{array}\right)$  terms, our algorithm works better in most cases.
%
For probabilistic algorithms,
our method is the only one whose complexity is linear in $nT$
and is better than that of the Monte Carlo method given in \cite{2}.

Kaltofen gave an interpolation algorithm for SLP polynomials \cite{5d}, whose complexity is polynomial in $D$. Avenda\~{n}o-Krick-Pacetti \cite{5c} gave an algorithm for interpolating an SLP $f\in \Z[x]$, with bit complexity polynomial in $L,\log D,h$ and $h'$, where $h$ is an upper bound on the height of $f$ and $h'$ an upper bound on the height of the values of $f$ (and some of its derivatives) at some sample points. This method does not seem to extend to arbitrary rings. Mansour \cite{5b} and Alon-Mansour \cite{5a} gave deterministic algorithms for polynomials in $\Z[x]$, with bit complexity polynomial in $n,\log D,T,H$, where $H$ is an upper bound on the bit-length of the output coefficients. Also, this method is hard to extend to arbitrary rings.
Note that interpolation algorithms for black-box polynomials \cite{3,813,13,9}
can also be used to SLP polynomials,
but the complexities of these algorithms are polynomial in $D$ instead of $\log D$.

\noindent
\begin{table}[ht]
\footnotesize
\centering
\begin{tabular}{c|c|c|c}
&Bit&Field &Algorithm\\
&Complexity &Extension&type \\ \cline{1-4}
Garg $\&$ Schost \cite{5}&$Ln^2T^4\log^2D\log q$&not&Deterministic\\
Randomized Garg-Schost \cite{8}&$Ln^2T^3\log^2D\log q$&not&Las Vegas\\
Giesbrecht $\&$ Roche \cite{8}&$Ln^2T^2\log^2 D(n\log D+\log q)$&yes&Las Vegas\\
Arnold, Giesbrecht $\&$ Roche \cite{2}&$Ln^3T\log^3 D\log q$&not&Monte Carlo\\
Arnold, Giesbrecht $\&$ Roche \cite{2i}&$LnT\log^2 D(\log D+\log q)+n^\omega T$&yes&Monte Carlo\\
Arnold, Giesbrecht $\&$ Roche \cite{2j}&$Ln\log D(T\log D+n)(\log D + \log q) $&yes&Monte Carlo\\
&$+ n^{\omega-1} T\log D+ n^{\omega}\log D$&&\\
%
\cline{1-4}
This paper (Thm \ref{the-2})&$L n^2T^2\log^2 D\log q+LnT\log^3 D\log q$&not&Deterministic\\
This paper (Thm \ref{them-5})&$LnT\log^3 D\log q$&not&Monte Carlo\\
This paper (Thm \ref{them-7}) &$LnT\log^2 D(\log q+\log D)$&yes&Monte Carlo
\end{tabular}
\caption{A ``soft-Oh" comparison  for SLP  polynomials over finite field $F_q$}\label{tab-2}
\end{table}

In Table \ref{tab-2}, we list the bit complexities for  SLP interpolation algorithms over the finite field $\F_q$. In the table,
 ``field extension"  means that in the probe of the SLP, whether elements in an extension field of $\F_q$ are needed.
It is easy to see that our probabilistic algorithms are the only ones
which are linear in $n$ and $T$.
Also, our second probabilistic algorithm has better complexity than all
algorithms except randomized Garg-Schost \cite{8} which is Las Vegas.
The ratio of the cost of our first probabilistic algorithm with that of
the randomized Garg-Schost \cite{8} is $\frac{\log D}{nT^2}$,
so our method is faster in most cases.

\subsection{Main idea and relation with existing work}
%
%
%
%
%

Our methods build on the work by Garg-Schost \cite{5},
Arnold-Giesbrecht-Roche \cite{8}, Giesbrecht-Roche \cite{6}, and Klivans-Spielman \cite{9}, where the basic idea is to reduce multivariate interpolation to univariate interpolation.
Three new techniques are introduced in this paper:
a criterion for checking whether a term belongs to a polynomial
(see Section \ref{sec-2} for details),
a deterministic method to find an ``ok" prime
(see Section \ref{sec-3} for details),
a new Kronecker type substitution to reduce  multivariate polynomial interpolation to univariate polynomial interpolation
(see Section \ref{sec-mi1} for details).
Our methods have three major steps:
\begin{itemize}
\item
First, we find an ``ok" prime $p$ such that at least half of the
terms of $f$ do not collide or merge with other terms of $f$ in
the univariate polynomial $f_{(D,p)}^{\modp} = f(x,x^D,\dots,x^{D^{n-1}})\ \modp\ (x^p-1)$.

\item
Second, we obtain a set $S$ of terms containing those non-colliding terms
of $f$. In the univariate case,  these terms are found by the Chinese Remaindering Theorem and in the multivariate case, these terms are found by a new Kronecker
substitution.

\item
Finally, we use our criterion for checking whether a term belongs to a polynomial
to find at least half of the terms of $f$ from $S$.
\end{itemize}
Repeating these three steps for at most $\log_2 T$ times, we obtain $f$.
In the rest of this section, we give detailed comparison with related work.

Grag and Schost \cite{5} gave a deterministic interpolation algorithm for a univariate SLP polynomial $f$  by recovering $f$ from $f\ \modp\ (x^{p}-1)$ for $\mathcal{O}(T^2\log D)$ different primes $p$.
The randomized Las Vegas version of this method needs $\mathcal{O}(T\log D)$ probes.
The multivariate interpolation
comes directly from the Kronecker substitution \cite{10}.
Our univariate interpolation algorithm has two major differences
from that given in \cite{5}.
First, we compute $f\ \modp\ (x^{p}-1)$ for $\mathcal{O}(T\log D)$
different primes $p$, and second we introduce a criterion to check whether a term really belongs to $f$.
Our multivariate interpolation method is similar to our univariate interpolation algorithm, where a new Kronecker type substitution is introduced to recover the exponents.

Giesbrecht-Roche \cite{8} introduced the idea of diversification
and a probabilistic method to choose  ``good" primes.
It improves Grag and Schost's algorithm by a factor $\mathcal{O}(T^2)$, but becomes a Las Vegas algorithm.
In our Monte Carlo algorithms, we use  a new Kronecker substitution instead of the method of diversification to find the same term in different remainders of $f$, where only additions are used for the coefficients. Hence, our algorithm can work for more general rings and has better complexity.
%

In Arnold, Giesbrecht, and Roche \cite{2}, the concept of
``ok" prime is introduced and a Monte Carlo univariate algorithm is given, which has complexity linear in $T$ but cubic in $n$.
%
The ``ok" prime in \cite{2} is probabilistic.
In our deterministic method, we give a method to find an exact ``ok" prime.
%

In Arnold, Giesbrecht, and Roche \cite{2i}, their univariate interpolation algorithm is extended to finite fields.
By combining the idea of diversification, the complexity becomes better. This algorithm will be used in our second probabilistic  algorithm.
The bit complexity of their multivariate interpolation
algorithm is linear in $n^\omega$, where $\omega$ is the constant of matrix multiplication, while our algorithm is linear in $n$.
The reason is that, our method uses a new Kronecker substitution to find
the exponents and does not need to solve linear systems.

In Arnold, Giesbrecht, and Roche \cite{2j}, they  further
improved their interpolation algorithm for finite fields. By combining the random Kronecker substitution and diversification, the complexity becomes better, but still linear in $n^\omega$.

Finally, the new Kroneceker type substitution introduced in this paper is inspired by the works of Klivans-Spielman \cite{9}, Arnold \cite{0} and Arnold-Roche \cite{1}.
In \cite{9}, the substitution $f(q_1x,q_2x^{\modp(D,p)},\\\dots,q_n x^{\modp(D^{n-1},p)})$ was used,  where $q_1,\ldots,q_n$ are primes.
In this paper, we introduced the substitution
 $f(x,x^{\modp(D,p)},$ $\dots,$ $x^{p+\modp(D^{k-1},p)},$ $\ldots,$ $x^{\modp(D^{n-1},p)})$ (see section \ref{sec-mi1} for exact definition).
Our substitution has the following advantages:
(1) For the complex filed, the size of data is not changed after our substitution,
while the size of data for the substitution in \cite{9} is increased by a factor of $D$.
(2) Only arithmetic operations for the coefficients are used in our algorithm and thus the algorithm works for general computable rings, while
the substitution in \cite{9} needs factorization and $\RB$
should be a UFD at least.
In \cite{0}, the substitution $f(x^{s_1},x^{s_2},x^{s_k+p}\dots,x^{s_n})$ was used,
where $s_i$ are random integers.
Comparing to the randomized Kronecker substitution in \cite{0,1}, our substitution is deterministic.

\section{A criterion for term testing}
\label{sec-2}

In this section, we give a criterion to check whether
a term belongs to a  polynomial.
%

Throughout this paper, let
$ f=c_1m_1+c_2m_2+\cdots+c_t m_t\in \RB[\X]$
be a multivariate polynomial with terms $c_im_i$,
where $\RB$ is a computable ring, $\X=\{x_1,x_2,\ldots,x_n\}$ are $n$ indeterminates, and $m_i,i=1,2,\dots,t$ are distinct monomials.
Denote $\#f=t$ to be the number of terms of $f$
and  $M_f=\{c_1m_1,c_2m_2,\dots,c_tm_t\}$ to be the set of terms of $f$.
Let  $D,T\in\N$ such that $D> \deg(f)$ and $ T\ge\#f$.
For $p\in\N_{>0}$, let
\begin{equation}
f^{\modp}_{(D,p)} = f(x,x^D,\dots,x^{D^{n-1}})\ \modp\ (x^p-1) \in\RB[x].
\end{equation}

We have the following key concept.
\begin{defi}\label{def-c1}
A term $cm\in M_f$  is called a collision in $f^{\modp}_{(D,p)}$ if there exists
an $aw\in M_f\backslash \{cm\}$ such that $m^{\modp}_{(D,p)}=w^{\modp}_{(D,p)}$.
%
\end{defi}

The following fact is obvious.
\begin{lemma}\label{lm-10}
Let $f\in \RB[x]$, $\deg (f)<D$ and $cm\in M_f$.
If $cm$ is not a collision in $f_{(D,p)}^{\modp}$, then for any prime $q$, $cm$ is also not a collision in $f^{\modp}_{(D,pq)}$.
\end{lemma}

\begin{lemma}\label{lm-1}
Let $f=\sum_{i=1}^t c_im_i$, $T\geq \# f,D>\deg f$, $N_1=\max\{1,\lceil n(T-1)\log D\rceil\}$. For each $cm\in M_f$, there exist at most $N_1-1$ primes $p_1,\ldots,p_{N_1-1}$ such that $cm$ is a collision in $f^{\modp}_{(D,p_i)}$ for all $i=1,\ldots,N_1-1$.
\end{lemma}
\proof
If $T=1$, then $N_1=1$. The lemma is obvious.
Now we assume $T\geq 2$, then $N_1=\lceil n(T-1)\log D\rceil$.
It suffices to show that for any $N_1$ different primes $p_1,p_2,\dots,p_{N_1}$,  there exists at least one $p_j$, such that $cm$ is not a collision in $f^{\modp}_{(D,p_j)}$.

Assume $m_i=x_1^{e_{i,1}}x_2^{e_{i,2}}\cdots x_n^{e_{i,n}},i=1,2,\dots,t$. We prove it by contradiction.
It suffices to consider the case of $c_1m_1$. We assume by contradiction that for every $p_j,j=1,2,\dots,N_1$, $c_1m_1$ is a collision in $f^{\modp}_{(D,p_j)}$.
Let
$$B=\prod^t_{s=2}(\sum_{i=1}^ne_{1,i}D^{i-1}-\sum_{i=1}^ne_{s,i}D^{i-1}).$$
First, we show that if $c_1m_1$ is a collision in $f^{\modp}_{(D,p_j)}$, then $\modp(B,p_j)=0$.
Since $c_1m_1$ is a collision in $f^{\modp}_{(D,p_j)}$, without loss of generality, assume $m^{\modp}_{1(D,p_j)}=m^{\modp}_{2(D,p_j)}$.
Then $0=\deg (m^{\modp}_{1(D,p_j)})-\deg (m^{\modp}_{2(D,p_j)})=\modp(\sum_{i=1}^ne_{1,i}D^{i-1},p_j)-\modp(\sum_{i=1}^ne_{2,i}D^{i-1},p_j)$.
So we have $\modp(\sum_{i=1}^ne_{1,i}D^{i-1}$ $-\sum_{i=1}^ne_{2,i}D^{i-1},p_j)=0$. So $\modp(B,p_j)=0$.

Since $p_1,p_2,\dots,p_{N_1}$ are different primes, $\prod^{N_1}_{j=1}p_j$ divides $B$.
Note that $|\sum_{i=1}^ne_{1,i}D^{i-1}-\sum_{i=1}^ne_{s,i}D^{i-1}|\leq (D-1)(\sum_{i=1}^nD^{i-1})=D^n-1$.
So $|B|=|\prod^t_{s=2}(\sum_{i=1}^ne_{1,i}D^{i-1}-\sum_{i=1}^ne_{s,i}D^{i-1})|\leq (D^n-1)^{t-1}$.
Thus $\prod^{N_1}_{j=1}p_j\geq 2^{N_1}\geq 2^{n(T-1)\log_2 D}=D^{n(T-1)}>(D^n-1)^{T-1}\geq |B|$, which  contradicts
the fact that $\prod^{N_1}_{j=1}p_j$ divides $B$.
The lemma is proved.\qed

Now we give a criterion for testing whether a term $cm$ is in $M_f$.
\begin{theorem}\label{lm-2}
Let $f=\sum_{i=1}^t c_im_i$, $T\geq \# f,D>\deg f$, $N_1=\max\{1,\lceil n(T-1)\log D\rceil\}$, $N_2=\lceil nT\log D\rceil$, and $\PP=\{p_1,p_2,\dots,p_{N_1+N_2-1}\}$ be $N_1+N_2-1$ different primes.
For a term $cm$ satisfying $\deg(m)<D$, $cm\in M_f$ if and only if there exist at least $N_2$
integers $j\in[1,N_1+N_2-1]$ such that $\#(f-cm)^{\modp}_{(D,p_j)}<\#f^{\modp}_{(D,p_j)}$.
\end{theorem}
\proof
If $T=1$, then $N_1=1$, the proof is obvious. So we assume $T\geq 2$, then $N_1=\lceil n(T-1)\log D\rceil$.

Let $cm\in M_f$. If $p_j$ is a prime such that $cm$ is not a collision in $f^{\modp}_{(D,p_j)}$, then $\#(f-cm)^{\modp}_{(D,p_j)}=\#f^{\modp}_{(D,p_j)}-1$. So $\#(f-cm)^{\modp}_{(D,p_j)}<\#f^{\modp}_{(D,p_j)}$.
By Lemma \ref{lm-1}, there exist at most $N_1-1$ primes $q_j$ such that $cm$ is a collision in $f^{\modp}_{(D,q_j)}$. In $\PP$, as $N_1+N_2-1-(N_1-1)=N_2$, there exist at least $N_2$ primes such that $\#(f-cm)^{\modp}_{(D,p_j)}<\#f^{\modp}_{(D,p_j)}$.

For the other direction, assume $cm\notin M_f$. We show there exist at most $N_2-1$ integers $j\in[1,N_1+N_2-1]$ such that $\#(f-cm)^{\modp}_{(D,p_j)}<\#f^{\modp}_{(D,p_j)}$.
Consider two cases:
Case 1: $m$ is not a monomial in $f$. Case 2: $m$ is a monomial in $f$, but $cm$ is not a term in $f$.

Case 1. Since $m$ is not a monomial in $f$, $-cm$ is a term in $M_{f-cm}$
and $\#(f-cm)\le T+1$.
By Lemma \ref{lm-1}, there exist at most $N_2-1$ primes in $\PP$ such that $-cm$ is a collision in $f-cm$. For all other primes $p_j$ in $\PP$, $\#((f-cm)-(-cm))^{\modp}_{(D,p_j)}=\#(f-cm)^{\modp}_{(D,p_j)}-1$, that is, $\#(f-cm)^{\modp}_{(D,p_j)}=\#f^{\modp}_{(D,p_j)}+1$.
So there exist at most $N_2-1$ primes $p_j$ in $\PP$ such that $\#(f-cm)^{\modp}_{(D,p_j)}<\#f^{\modp}_{(D,p_j)}$.

Case 2. Since $m$ is a monomial in $f$ and $cm\notin M_f$,  $f-cm$ has the same number of terms as $f$. Assume the term of $f$ with monomial $m$ is $c_1m$. Then $(c_1-c)m\in M_{f-cm}$. By Lemma \ref{lm-1},  for at most $N_1-1(\leq N_2-1)$ primes $p_j$ in $\PP$, $(c_1-c)m$ is a collision in $(f-cm)^{\modp}_{(D,p_j)}$. For all other primes $p_k$ in $\PP$, $(c_1-c)m$ is not a collision in $(f-cm)^{\modp}_{(D,p_k)}$,
or equivalently,  $\#(f-cm)^{\modp}_{(D,p_k)}=\#(f-cm-(c_1-c)m)^{\modp}_{(D,p_k)}+1 =\#(f-c_1m)^{\modp}_{(D,p_k)}+1$. But we always have $\#(f-c_1m)^{\modp}_{(D,p_k)}+1\ge\#f^{\modp}_{(D,p_k)}$, and so
$\#(f-cm)^{\modp}_{(D,p_k)}\ge\#(f)^{\modp}_{(D,p_k)}$.
So there exist at most $N_2-1$ primes $p_j$ in $\PP$ such that $\#(f-cm)^{\modp}_{(D,p_j)}<\#f^{\modp}_{(D,p_j)}$.
The theorem is proved.\qed

As a corollary, we can deterministically recover $f$ from $f^{\modp}_{(D,p_j)},j=1,2,\dots,N_1+N_2-1$.
\begin{cor}\label{lm-4}
Use the notations in Theorem \ref{lm-2}.
We can uniquely recover $f$ from $f^{\modp}_{(D,p_j)},j=1,2,\dots,N_1+N_2-1$.
\end{cor}
\proof
Let $C=\{c_1,c_2,\dots,c_k\}$ be all the different coefficients in $f^{\modp}_{(D,p_j)},j=1,2,\dots,N_1+N_2-1$. By Lemma \ref{lm-1}, since $N_1+N_2-1\geq N_1$, all the coefficients of $f$ are in $C$. Let $M=\{m_1,m_2,\dots,m_s\}$ be the set of all the monomials with degrees less than $D$.
So all the terms of $f$ are in $\{c_im_j|i=1,2,\dots,k,j=1,2,\dots,s\}$. By Theorem \ref{lm-2}, we can check if $c_im_j$ is in $M_f$. So we can find all the terms of $f$.\qed

The above result can be changed into a deterministic algorithm for interpolating $f$.
But the algorithm is not efficient due to the reason that
$s$ is linear in $D^n$.
In the following, we will show how to find a smaller alternative set $M$ and give an efficient interpolation algorithm.

\section{Find an ``ok" prime}
\label{sec-3}
A prime $p$ is called an {\em ``ok" prime} if at least half of the terms in
$f$ are not collisions in $f^{\modp}_{(D,p)}$, where $D>\deg f$.
In this section, we give a deterministic method to find an ``ok" prime
for $f$.

Denote $\mathcal{C}^f_{(D,p)}$ to be the number of collision terms of $f$ in $f^{\modp}_{(D,p)}$.
We need the following lemma from \cite{2}.

\begin{lemma}\label{lm-3}\cite{2}
Let $f\in \RB[x]$. If $\#f^{\modp}_{(D,q)}\le \#f^{\modp}_{(D,p)}$, then $\mathcal{C}^f_{(D,p)}\leq 2\mathcal{C}^f_{(D,q)}$.
\end{lemma}

It is easy to modify the above lemma into multivariate case.
\begin{cor}\label{cor-1}
 Let $f\in\RB[\X]$, $D>\deg(f)$. If $\#f^{\modp}_{(D,q)}\leq\#f^{\modp}_{(D,p)}$, then $\mathcal{C}^f_{(D,p)}\leq 2\mathcal{C}^f_{(D,q)}$.
\end{cor}
\proof
Let $F=f(x,x^D,\dots,x^{D^{n-1}})$, then $f^{\modp}_{(D,p)}=F\ \modp\ (x^p-1)$. So $\mathcal{C}^f_{(D,p)}=\mathcal{C}^F_{(D,p)}$. The corollary follows from Lemma \ref{lm-3}.\qed

%
\begin{lemma}\label{lm-6}
Let $f=\sum_{i=1}^tc_im_i\in\RB[\X]$, $m_i=x_1^{e_{i,1}}x_2^{e_{i,2}}\cdots x_n^{e_{i,n}}$, $D>\deg(f)$, $A=\prod_{i,j\in\{1,\dots,t\}}^{i< j}\sum_{k=1}^n\\(e_{i,k}-e_{j,k})D^{k-1}$,
and $p$ a prime. If $\mathcal{C}^f_{(D,p)}=s$, then $p^{\lceil\frac{s}{2}\rceil}$
divides $A$.
\end{lemma}
\proof
We divide the terms of $f$ into groups, called {\em collision blocks}, such that
two terms of $f$ collide  in $f^{\modp}_{(D,p)}$
if and only if they are in the same  group.
Let $n_i$ be the number of collision blocks containing $i$ terms.
Assume $c_{j_1}m_{j_1}+c_{j_2}m_{j_2}+\cdots+c_{j_i}m_{j_i}$ is in a collision block with $i$ terms. For any $u,v\in\{j_1,j_2,\dots,j_i\}$, we have $\deg (m^{\modp}_{u(D,p)})=\deg (m^{\modp}_{v(D,p)})$.
So $(e_{u,1}+e_{u,2}D+\cdots+e_{u,n}D^{n-1})\ \modp\ p=(e_{v,1}+e_{v,2}D+\cdots+e_{v,n}D^{n-1})\ \modp\ p$, which implies that $p$ divides $(e_{u,1}-e_{v,1})+(e_{u,2}-e_{v,2})D+\ldots+(e_{u,n}-e_{v,n})D^{n-1}$.

There are $C_i^2=\frac{i(i-1)}{2}$ pairs such $u,v$, so $p^{\frac{i(i-1)}{2}}$ is a factor of $A$. Let $K=\sum_{i=1}^t\frac{1}{2}(i^2-i)n_i$. Since there exist $n_i$ such collision blocks, $p^K$ is a factor of $A$.
Now we give a lower bound of $K$. First we see that $t=\sum_{i=1}^tin_i,s=\sum_{i=2}^tin_i$.
$K=\sum_{i=1}^t\frac{1}{2}(i^2-i)n_i=\frac{1}{2}\sum_{i=1}^ti^2n_i-\frac{1}{2}\sum_{i=1}^tin_i=\frac{1}{2}\sum_{i=1}^ti^2n_i-\frac{1}{2}t=\frac12n_1+\frac{1}{2}\sum_{i=2}^ti^2n_i-\frac{1}{2}t\geq
\frac12n_1+t-n_1-\frac12t=\frac12 t-\frac12n_1=\frac12 s$.
If $n_i=0,i\geq 3$, then $K=\frac12 s=\lceil\frac12 s\rceil$.
If there is at least one $n_i>0,i>3$, then $K>\frac12 s$. So $K\geq \lceil\frac12 s\rceil$. We have proved the lemma.\qed

\begin{theorem}\label{lm-21}
Let $f=\sum_{i=1}^tc_im_i\in\RB[\X]$, $T\geq \# f,D>\deg f$, $N_1=\max\{1,\lceil n(T-1)\log D\rceil\}$, and $p_1,p_2,\dots,p_{4N_1}$ be $4N_1$ different primes.
Let $j_0$ be an integer in $[1,4N_1]$ such that $\# f^{\modp}_{(D,p_{j_0})}\geq \# f^{\modp}_{(D,p_{j})}$ for all $j$.
%
Then at least $\lceil\frac t 2\rceil$ of the terms of $f$  are not collisions in $f^{\modp}_{(D,p_{j_0})}$.
\end{theorem}
\proof
If $T=1$, then $N_1=1$, the proof is obvious. So we assume $T\geq 2$, then $N_1=\lceil n(T-1)\log D\rceil$.
We first claim that there exists at least one $p_j$ in $p_1,p_2,\dots,p_{4N_1}$ such that $\mathcal{C}_{(D,p_j)}^f< \frac t4$. We prove it by contradiction. Assume for $j=1,2,\dots,4N_1$, $\mathcal{C}_{(D,p_j)}^f\geq \frac t4 $. Then by Lemma \ref{lm-6}, {for all $j$ in $[1,4N_1]$,} $p_j^{\lceil\frac12 \mathcal{C}_{(D,p_j)}^f\rceil}$ divides $A$,  where $A$ is defined in Lemma \ref{lm-6}. Since $p_j,j=1,2,\dots,4N_1$ are different primes, then $\prod_{j=1}^{4N_1}p_j^{\lceil\frac12\mathcal{C}_{(D,p_j)}^f\rceil}$ divides $A$.
Now $\prod_{j=1}^{4N_1}p_j^{\lceil\frac12 \mathcal{C}_{(D,p_j)}^f\rceil}\geq\prod_{j=1}^{4N_1}p_j^{\lceil\frac12 \cdot\frac14t\rceil}\geq 2^{4N_1\frac t 8}\geq 2^{\frac12tN_1}\geq 2^{\frac12 tn(T-1)\log_2 D}=D^{\frac12 nt(T-1)}$, which contradicts to the inequality $A\leq (D^n-1)^{\frac{t(t-1)}{2}}$. We proved the claim.

By Corollary \ref{cor-1}, we have $\mathcal{C}_{(D,p_{j_0})}^f\leq 2\mathcal{C}^f_{(D,p_{j})},j=1,2,\dots,4N_1$. So $\mathcal{C}_{(D,p_{j_0})}^f<2\cdot\frac t 4=\frac12 t$.
%
So the number of no collision terms of $f$ in $f^{\modp}_{(D,p_{j_0})}$ is $>t-\frac t 2=\frac12 t$. We proved the theorem.\qed

\section{Deterministic univariate interpolation}
In this section, we consider the interpolation of a univariate polynomial $f$ with $\deg f < D$.
The algorithm works as follows.
First, we use Theorem \ref{lm-21} to find an ``ok" prime $p$
such that at least half of the
terms of $f$ are not collisions in $f_{(D,p)}^{\modp}$.
Second, we use $f_{(D,pp_k)}^{\modp},k=1,2,\dots,K_D$
to find a set $S$  containing these non-collision terms of $f$ by the Chinese Remaindering Theorem, where $p_k$ is the $k$-th prime and $K_D$ is the smallest number such that $p_1p_2\dots p_{K_D}\geq D$.
Finally, we use Theorem \ref{lm-2} to pick up the terms of $f$ from $S$.

\subsection{Recovering terms from module $x^p-1$}

In this section, let $f$ be a  univariate polynomial in $\RB[x]$.
We will give an algorithm to recover those terms of $f$ from $f^{\modp}_{(D,p)}$, which are not collisions in $f^{\modp}_{(D,p)}$.

Let $f$ be a univariate polynomial, $D> \deg f$, and $p\in \N{>0}$. In this case, $f^{\modp}_{(D,p)}=f(x)\ \modp \ (x^p-1)$.
Write
\begin{eqnarray}
&&f_{(D,p)}^{\modp}=a_1x^{d_1}+a_2x^{d_2}+\cdots+a_r x^{d_r}\label{eq-ufdp}\\
&&f_{(D,pp_k)}^{\modp}=f_{k,1}+f_{k,2}+\cdots+f_{k,r}+g_k,\quad k=1,2,\dots,K_D\nonumber
\end{eqnarray}
where $d_1<d_2< \cdots < d_r$, $p_k$ is the $k$-th prime, $K_D$ is the smallest number such that $p_1p_2\dots p_{K_D}\geq D$,
and $f_{k,i}\ \modp\ (x^p-1)=a_ix^{d_i}$ and $g_k\ \modp\ (x^p-1)=0$.
$f_{(D,pp_k)}^{\modp}$ can be written as the above form,
because $f_{(D,pp_k)}^{\modp}\, \modp\, (x^p-1) = f_{(D,p)}^{\modp}$.
We now introduce the following key notation
\begin{eqnarray}
&&U^f_{D,p} =\{a_i x^{e_i}| \hbox{ such that } i\in[1,r],
a_i \hbox{ is from \bref{eq-ufdp}, }
e_i\in[0,D-1],\hbox{ and } \cr
 &&\quad \hbox{U1}: f_{k,i}=a_ix^{b_{k,i}},k=1,2,\dots,K_D.\label{eq-uf}\\ 
 &&\quad \hbox{U2}: \modp(e_i,p_k)=\modp(b_{k,i},p_k),k=1,2,\dots,K_D.\}\nonumber
\end{eqnarray}

The following lemma gives the geometric meaning of $U^f_{D,p}$.
\begin{lemma}\label{lm-13}
Let $f\in \RB[x]$, $\deg (f)<D$ and $cm\in M_f$.
If $cm$ is not a collision in $f_{(D,p)}^ {\modp}$, then $cm\in U^f_{D,p}$.
\end{lemma}
\proof
It suffices to show that $cm$ satisfies the conditions of the definition of $U^f_{D,p}$. Assume $m=x^e$.
Since $cm$ is not a collision in $f_{(D,p)}^{\modp}$,
without loss of generality, assume $cm_{(D,p)}^{\modp} = a_1x^{d_1}$
and $d_1 = \modp(e,p)$, where $a_1x^{d_1}$ is defined in \bref{eq-ufdp}.
By Lemma \ref{lm-10}, $cm$ is also not a collision in $f^{\modp}_{(D,pp_k)}$
and $f_{k,1} = a_1 x^{b_{k,1}}$ for $b_{k,1}=\modp(e,pp_k)$, $k=1,2,\dots,K_D$.
Since $\modp(e,p_k)=\modp(\modp(e,pp_k),p_k)=\modp(b_{k,1},p_k)$, conditions U1 and U2 are satisfied and the lemma is proved.\qed
%
%

Note that $U^f_{D,p}$ may contain terms not in $M_f$.   
The following algorithm computes the set $U^f_{D,p}$.
\begin{alg}[UTerms]\label{alg-1}
\end{alg}

{\noindent\bf Input:}
Univariate polynomials $f^{\modp}_{(D,p)},f^{\modp}_{(D,pp_k)}$, $k=1,2,\dots,K_D$, a prime $p$, a degree bound $D> \deg (f)$.

{\noindent\bf Output:} $U^f_{D,p}$.

\begin{description}
\item[Step 1:]
Write $f_{(D,p)}^{\modp}$ and $f^{\modp}_{(D,pp_k)}$ in as form  \bref{eq-ufdp}:
%
\begin{eqnarray*}
&&f_{(D,p)}^{\modp}=a_1x^{d_1}+a_2x^{d_2}+\cdots+a_rx^{d_r}\cr
&&f^{\modp}_{(D,pp_k)}=f_{k,1}+f_{k,2}+\cdots+f_{k,r}+g_k
\end{eqnarray*}
where $k=1,\dots,K_D$.

\item[Step 2:] Let $U=\{\}$.
For $i=1,2,\dots,r$
    \begin{description}
    \item[a:]
    $\mathbf{If}$ for $k=1,2,\dots,K_D$, one of $\#(f_{k,i})\neq 1$ or one of the coefficient of $f_{k,i}$ is not $a_i$, $\mathbf{then}$ break.
    \item[b:]
    for $k=1,2,\dots,K_D$, assume $f_{k,i}=a_ix^{b_{k,i}}$ and
 let $e_k=\modp(b_{k,i},p_k)$.

    \item[c:]Let $\beta=\mathbf{Chinese\ Remainder}([e_1,e_2,\dots,e_{K_D}],[p_1,p_2,\dots,p_{K_D}])$.

    \item[d:]    If $\beta< D$ then let $U=U\bigcup\{a_ix^{\beta}\}$.
    \end{description}

\item[Step 3:] Return $U$.
\end{description}

\begin{remark}
In $\mathbf{c}$ of Step 2, $\mathbf{Chinese\ Remainder}([e_1,e_2,\dots,e_{K_D}],[p_1,p_2,\dots,p_{K_D}])$ means to find an integer $0\leq w<D$ such that $\modp(w,p_i)=e_i,i=1,2,\dots,K_D$. Since $p_1p_2\cdots p_{K_D}\geq D$, the integer $w$ is unique. This can be done by the Chinese remainder algorithm.
\end{remark}

\begin{lemma}\label{lm-c1}
Algorithm \ref{alg-1} needs $O(T\log D)$ arithmetic operations in $\RB$ and $O^\thicksim(T\log D\log p+T\log^2 D)$ bit operations.
\end{lemma}
\proof
In Step 1, we need to do a traversal for the terms of $f^{\modp}_{(D,pp_k)}$. Since $K_D$ is the smallest number such that $p_1p_2\cdots p_{K_D}\geq D$, then $2^{K_D-1}\leq p_1p_2\cdots p_{K_D-1}<D$. So $K_D\leq\log_2 D+1$, which is $O(\log D)$.
In order to write $f^{\modp}_{(D,pp_k)}$ as the form \bref{eq-ufdp}, we needs to perform the modular operation $\modp\ \ p$ on every degree of $f^{\modp}_{(D,pp_k)}$, then use the quick sorting method to write their terms in ascending order according to the degree.
Since  $f^{\modp}_{(D,pp_k)}$ has no more than $T$ terms and $k=1,2,\dots,K_D$, it needs $O^\thicksim(T\log D)$ arithmetic operations. Since the height of the data is $O(\log (pp_{K_D}))$ and the prime $p_{K_D}$ is $O^\thicksim(\log D)$, it needs $O^\thicksim(T\log D\log (pp_{K_D}))$ bit operations, which is $O^\thicksim(T\log D\log p+T\log^2 D)$ bit operations.

In $\mathbf{a}$ of Step 2, since $\#f^{\modp}_{(D,pp_k)}\leq T$, it totally needs $O^\thicksim(T\log D)$ bit operations to determine whether $\#(f_{k,i})=1$. To compare the coefficients of $f_{k,i}$, it needs $O(T\log D)$ arithmetic operations in $\RB$.
In $\mathbf{b}$ of Step 2, since the height of the data is $O(\log (pp_{K_D}))$, it needs $O(T\log^2 D+T\log D\log p)$ bit operations. In $\mathbf{c}$, we need to call at most $T$ times Chinese remaindering. By \cite[p.290]{12}, the cost of the Chinese remaindering algorithm is $O^\thicksim(\log D )$ arithmetic operations in $\Z$. Since the height of the data is $O(\log D)$, it needs $O^\thicksim(T\log^2 D)$ bit operations. So the complexity of Step 2 is $O^\thicksim(T\log^2 D+T\log D \log p)$ bit operations and $O(T\log D)$ arithmetic operations.
\qed

\subsection{Interpolation algorithm for univariate polynomials}

We first give a precise definition for SLP polynomials.
\begin{defi}\label{def-slp}
An SLP over a ring $\RB$ is a branchless sequence of arithmetic instructions that represents a polynomial function. It takes as input a vector $(a_1,\dots,a_n)$ and outputs a vector $(b_1,\dots,b_L)$ by way of a series of instructions $\Gamma_i:1\leq i\leq L$ of the form $\Gamma_i:b_i\leftarrow \alpha\star_i \beta$, where $\star_i$ is an operation $'+','-'$ or $'\times'$, and $\alpha,\beta\in\RB\bigcup\{a_1,\dots,a_n\}\bigcup\{b_0,\dots,b_{i-1}\}$.
The inputs and outputs may belong to $\RB$ or a ring extension of $\RB$. We say that an SLP computes a multivariate polynomial $f\in\RB[x_1,\dots,x_n]$ if it sets $b_L$ to be $f(a_1,\dots,a_n)$.
\end{defi}

We now give the interpolation algorithm for univariate polynomials,
where $f^*$ in the input is introduced because the multivariate
interpolation algorithm in Section \ref{sec-mi} will use it.

\begin{alg}[UIPoly]\label{alg-2}
\end{alg}

{\noindent\bf Input:} An  SLP $\S_f$ that computes $f(x)\in\RB[x]$, $f^*\in\RB[x]$, $T\geq \max\{\# f,\#f^*\},T_1\geq\#(f-f^*)$, $(T\geq T_1)$, $D>\max\{\deg f,\deg f^*\}$.

{\noindent\bf Output:} The exact form of $f-f^*$.

\begin{description}
\item[Step 1:] Let $N_1=\max\{1,\lceil (T_1-1)\log_2 D\rceil\},N_2=\lceil T_1\log_2 D\rceil,N=\max\{4N_1,N_1+N_2-1\}$.

\item[Step 2:] Find the first $N$ primes $p_1, \dots,p_N$.
\item[Step 3:]
Compute the smallest  $K_D$ such that $p_1 \cdots p_{K_D}\geq D$.

\item[Step 4:]
For $j=1,2,\dots,N$, probe
$f^{\modp}_{(D,p_j)}$.
Let $f_j=f^{\modp}_{(D,p_j)}-f^{*\modp}_{(D,p_j)}$ and $h^*=0$.

\item[Step 5:] Loop
\begin{description}
\item[5.1:] Let $\alpha=\max\{\#f_j|j=1,2,\dots,N\}$ and $j_0$  the smallest number such that $\#f_{j_0}=\alpha$.

\item[5.2:] If $\alpha=0$, then return $h^*$.

\item[5.3:] For $k=1,2,\dots,K_D$, probe $f^{\modp}_{(D,p_{j_0}p_k)}$ and let $g_k=f^{\modp}_{(D,p_{j_0}p_k)}-f^{*\modp}_{(D,p_{j_0}p_k)}$.

\item[5.4:] Let $U^{f-f^*-h^*}_{D,p_{j_0}}:=\mathbf{UTerms}(f_{j_0},g_1,g_2,\dots,g_{K_D},p_{j_0},D)$.

\item[5.5:] Let $h=0$. For each $u\in U^{f-f^*-h^*}_{D,p_{j_0}}$,
if
$$\#\{j\,|\, \#(f_j-u^{\modp}_{(D,p_j)})<\#(f_j),j=1,\dots,N_1+N_2-1 \}\ge N_2$$
then $h:=h + u$.

\item[5.6:] Let $h^*=h^*+h$, $T_1=T_1-\#h$, $N_1=\max\{1,\lceil (T_1-1)\log_2 D\rceil\},N_2=\lceil T_1\log_2 D\rceil, N=\max\{4N_1,N_1+N_2-1\}$.

\item[5.7:] For $j=1,2,\dots,N$, let $f_j=f_j-h^{\modp}_{(D,p_{j})}$.

\end{description}


\end{description}

%

\begin{theorem}\label{the-1}
Algorithm \ref{alg-2} returns $f-f^*$ using $O^\thicksim(LT^2\log^2 D+LT\log^3 D)$ ring operations in $\RB$
and similarly many bit operations, where $L$ is the size of the SLP representation for $f$.
Specially, when  $f^*=0$, the algorithm returns $f$.
\end{theorem}
\proof
We first prove the correctness of the theorem.
We claim that each loop of Step 5 will obtain at least half of the terms of $f-f^*-h^*$.
Then, the algorithm will return the correct $f-f^*$ by running at most
$\log_2 T_1$ times of the loop in Step 5.
In Step 5.1, Theorem \ref{lm-21} is used to find an okay prime $p_{j_0}$.
In Step 5.4,  by Lemma \ref{lm-13} and Theorem \ref{lm-21}, at least half of the terms of $f-f^*-h^*$ are in $U^{f-f^*-h^*}_{D,p_{j_0}}$.
In Step 5.5, Theorem \ref{lm-2} is used to select the elements of $M_{f-f^*-h^*}$ from $U^{f-f^*-h^*}_{D,p_{j_0}}$.
In summary, at Step 5.6,  $h$ contains at least half of the terms of $f-f^*-h^*$ and the claim is proved.
Then, the correctness of the algorithm is proved.

We now analyse the complexity of the algorithm, which comes from
Step 4 and Step 5. The complexity of other steps are lower
than these two steps.

In Step 3, since the bit complexity of finding the first $N$ primes is $O(N\log^2N\log\log N)$ by \cite[p.500,Thm.18.10]{12} and $N$ is $O^\thicksim(T\log D)$, the bit complexity of Step 3 is $O^\thicksim(T\log D)$.

In Step 4, we probe $N$ univariate polynomials $f^{\modp}_{(D,p_j)}$
and probing $f^{\modp}_{(D,p_j)}$ costs $O^\thicksim(Lp_j)$,
 since $S_f$ is of length $L$ and the univariate polynomials in the procedure have degrees $< p_j$.
Since  $p_i$ is of $O^\thicksim(T\log D)$ and $N$ is $O^\thicksim(T\log D)$, the cost of probing $f^{\modp}_{(D,p_j)}$ is  $O^\thicksim(LT^2\log^2 D)$ ring and bit operations. It needs $O^\thicksim(T^2\log D)$ ring operations and $O^\thicksim(T^2\log^2 D)$ bit operations to obtain $f^{*\modp}_{(D,p_{j})}$.
Then the total complexity of Step 4 is $O^\thicksim(LT^2\log^2 D)$ ring and bit operations.

We now consider Step 5. We first consider the complexity of each loop of the this step.
In Step 5.1, since $N$ is of $\mathcal{O}(T\log D)$ and the terms of $(f-f^*-h^*)^{\modp}_{(D,p_j)}$ is no more than $T$, it needs $O^\thicksim(T^2\log D)$ bit operations.

In Step 5.3, since $K_D$ is of $O(\log D)$ and $p_{j_0}p_k$ is of $O^\thicksim(T\log^2 D)$, we need  $O^\thicksim(LT\log^3 D)$ arithmetic operations in $\RB$ and similarly many bit operations to obtain $f^{\modp}_{(D,p_{j_0}p_k)}$.
We need $O(K_D\#f^*)$ ring operations and $O^\thicksim(K_D\#f^*\max\{\log \deg f^*,\log (p_{j_0}p_k)\})$ bit operations to obtain $f^{*\modp}_{(D,p_{j_0}p_k)}$.
Since $\#f^* \le T$, $\deg f^* \le D$, $p_k$, $K_D$ is of $O(\log D)$, and $p_{j_0}$ is of $O(T\log D)$, the cost is $O(T\log D)$ ring operations and $O^\thicksim(T\log D\max\{\log D, \log T+\log\log D\}) = O^\thicksim(T\log^2 D)$ bit operations.
Then, the total complexity of this step is $O^\thicksim(LT\log^3 D)$ arithmetic operations in $\RB$ and similarly many bit operations.

In Step 5.4, by Lemma \ref{lm-c1}, the complexity is $O(T\log D)$ arithmetic operations in $\RB$ and  $O^\thicksim(T\log^2 D)$ bit operations.

In Step $5.5$, in order to determine whether $\#(f_j-u^{\modp}_{(D,p_j)})<\#(f_j)$,
we just need to determine whether $u^{\modp}_{(D,p_j)}$ is a term of $f_j$.
We sort the terms of $f_j$ such that they are in ascending order according to their degrees, which costs $O^\thicksim((N_1+N_2) T)=O^\thicksim(T^2 \log D)$ bit operations,
since $N_1+N_2$ is $O^\thicksim(T\log D)$.
To find whether $f_j$ has a term with degree $\deg(u^{\modp}_{(D,p_j)})$, we need $O(\log T)$ comparisons. Since the height of the degree is $O(\log D)$, it needs $O(\log T\log D)$ bit operations. To compare the coefficient, it needs one arithmetic operation. So it totally needs $O(\log T\log D)$ bit operations and $O(1)$ arithmetic operation to compare $\#(f_j-u^{\modp}_{(D,p_j)})$ with $\#(f_j)$.
Hence, the total complexity of Step 5.5 is  $O^\thicksim(\#U^{f-f^*}_{D,p_{j_0}}
(N_1+N_2)\log T\log D + (N_1+N_2) T)$
bit operations and $O(\#U^{f-f^*}_{D,p_{j_0}}
(N_1+N_2))$ arithmetic operations in $\RB$.
Since $\#U^{f-f^*}_{D,p_{j_0}}\leq T$ and $N_1+N_2-1$ is of $O(T\log D)$, the total complexity is $O^\thicksim(T^2\log D)$ arithmetic operations and $O^\thicksim(T^2\log^2 D)$ bit operations.
%
%

So the total complexity of each loop of Step 5 is
$O^\thicksim(LT\log^3 D + T^2\log D)$ arithmetic operations and
$O^\thicksim(LT\log^3 D+T^2\log^2 D)$ bit operations, which comes from Step 5.3 and Step 5.5, respectively.
Since each loop of Step 5 will obtain at least   half of $f-f^*-h^*$,
Step 5 has at most $O(\log T)$ loops.
So the total complexity of Step 5
is $O^\thicksim(LT\log^3 D+T^2\log D)$ ring operations and $O^\thicksim(LT\log^3 D+T^2\log^2 D)$ bit operations.

Combing with the complexity of Step 4,
the total complexity of the algorithm is 
$O^\thicksim(LT^2\log^2 D+ LT\log^3 D+T^2\log^2 D) = O^\thicksim(LT^2\log^2 D+ LT\log^3 D)$ ring and bit operations, which
comes from Step 4 and Step 5.3. The theorem is proved.
\qed

For an $n$-variate polynomial $f$ of degree $<D$, we can use the Kronecker substitution \cite{10} to reduce the interpolation of $f$ to that of a univariate polynomial of degree $D^n$, which
can be computed with Algorithm $\mathbf{UIPoly}(S_f,0,T,T,D)$.
By Theorem \ref{the-1}, we  have
\begin{cor}\label{th-m1}
For an SLP $f\in\RB[\X]$ with $T\geq \#f$ and $D<\deg(f)$, we can find  $f$ using   $O^\thicksim(Ln^2T^2\log^2 D+Ln^3T\log^3 D)$ ring operations in $\RB$ and a similar number of bit operations.
\end{cor}
%

In the next section, we will give a multivariate polynomial interpolation algorithm
which has better complexity.

\section{Deterministic multivariate polynomial interpolation}
\label{sec-mi}
In this section, we will give a new multivariate interpolation  algorithm
which is quadratic in $n$, while the algorithm given in Corollary \ref{th-m1}
is cubic in $n$.
The algorithm is quite similar to Algorithm \ref{alg-2} and works as follows.
First, we use Theorem \ref{lm-21} to find an ``ok" prime $p$ for $f$.
Second, we use a modified Kronecker substitution to obtain a set $S$ of terms,
which contains at leats half of the terms of $f$.
Finally, we use Theorem \ref{lm-2} to identify the terms of $f$ from $S$.
The multivariate interpolation  algorithm will call Algorithm \ref{alg-2}.

\subsection{Recovering terms from module $x^p-1$}
\label{sec-mi1}

Let $f$ be a  multivariate polynomial,
$M_f$ the set of terms in $f$, $t=\#f$, $D> \deg(f)$, $T\ge\#f$, and $p\in\N_{>0}$.
Consider the modified Kronecker substitutions:
\begin{eqnarray}
f_{(D,p)} &=& f(x,x^{\modp(D,p)},\dots,x^{\modp(D^{n-1},p)}) \label{eq-f11}\\
f_{(D,p,k)} &=&
f(x,x^{\modp(D,p)},\dots,x^{p+\modp(D^{k-1},p)},\dots,x^{\modp(D^{n-1},p)})
\label{eq-f12}
\end{eqnarray}
where $k\in\{1,2,\dots,n\}$, $f_{(D,p)}$ comes from the substitutions
$x_i=x^{\modp(D^{i-1},p)},i=1,2,\dots,n$, and
$f_{(D,p,k)}$ comes from the substitutions $x_i= x^{\modp(D^{i-1},p)},i=1,2,\dots,n,i\neq k,x_k= x^{p+\modp(D^{k-1},p)}$.
%
Note that when $n=1$, $f_{(D,p)}=f_{(D,p,k)}=f(x)$.
Substitution \bref{eq-f11} was introduced in  \cite{9}
and substitution \bref{eq-f12} is introduced in this paper.
We have
\begin{equation}
\deg f_{(D,p)} \le Dp\hbox{ and }\deg f_{(D,p,k)} \le 2Dp.
\end{equation}
 
Similar to Definition \ref{def-c1},
a term $cm$ is said to be a {\em collision} in $f_{(D,p)}$ or in $f_{(D,p,k)}$,
if  there exists an $aw\in M_f\backslash \{cm\}$ such that $m_{(D,p)}=w_{(D,p)}$ or $m_{(D,p,k)}=w_{(D,p,k)}$.

We now show how to compute  $\S_{f_{(D,p)}}$ and $\S_{f_{(D,p,k)}}$.
\begin{lemma}\label{lm-pr1}
Let $\S_f$ be an SLP procedure to compute $f$, which has length $L$.
Then we can design a procedure $\S_{f_{(D,p)}}$ ($\S_{f_{(D,p,k)}}$ ) for $f_{(D,p)}$ (${f_{(D,p,k)}}$), which has length $L$ and costs extra $O(\log D+n\log p+L\log p)$ bit operations. Probing $f_{(D,p)}\ \modp \ (x^q-1)$ from $\S_{f_{(D,p)}}$
costs $O^\thicksim(Lq+L\log p)$ arithmetic operations and similarly many bit operations.
\end{lemma}
\proof
Define a procedure $\S_{f_{(D,p)}}$ for $f_{(D,p)}$ as follows.
Suppose we want to compute $f_{(D,p)}(a)$ for some $a$ in $\RB$
or an extension of $\RB$.
Assume $\S_f$ consists of the operations
$\Gamma_i:b_i\leftarrow \alpha_i \star_i \beta_i ,i=1,2,\dots,L$
with input $\{a_1,a_2,\dots,a_n\}$.
Now we define the $i$-th instruction $\overline{\Gamma}_i$ in $\S_{f_{(D,p)}}$. %
\begin{equation}
\overline{\Gamma}_i:=\begin{cases}
b_i\leftarrow \alpha_i\star_i \beta_i &\text{if both }\ \alpha_i\ \text{and } \beta_i\ \text{are not in } \{a_1,a_2,\dots,a_n\} \\
b_i\leftarrow a^{\modp(D^{j-1},p)}\star_i \beta_i &\text{if }\ \alpha_i\ \text{is}\ a_j, \text{but}\ \beta_i\ \text{is not in } \{a_1,a_2,\dots,a_n\} \\
b_i\leftarrow \alpha_i\star_i a^{\modp(D^{j-1},p)} &\text{if }\ \alpha_i\ \text{is not in}\ \{a_1,a_2,\dots,a_n\}, \text{but}\ \beta_i\ \text{is } a_j \\
b_i\leftarrow a^{\modp(D^{j_1-1},p)}\star_i a^{\modp(D^{j_2-1},p)} &\text{if }\ \alpha_i\ \text{is}\ a_{j_1} \ \text{and}\ \beta_i\ \text{is } a_{j_2} \\
\end{cases}
\end{equation}

Now we analyse the complexity of the procedure. In order to obtain $\modp(D^{j-1},p),j=1,2,\dots,n$, it needs $O(\log D+n\log p)$ bit operations. To obtain all $\Gamma_i$, it needs $O(L)$ arithmetic operations. Since the height of the data is $\log p$, it needs $O(L\log p)$ bit operations. So it totally needs $O(\log D+n\log p+L\log p)$ bit operations.

The univariate polynomial
$f_{(D,p)}\ \modp\ (x^q-1)$ can be computed from $\S_{f_{(D,p)}}$ as follow: first we replace $a_i$ by $x$.
During the computing, we always use the $\modp\ (x^q-1)$ to reduce the degree. So the degree of $x$ is less than $q$. If the length of $S_f$ is $L$, then probing $f_{(D,p)}\ \modp\ (x^q-1)$ from $\S_{f_{(D,p)}}$ costs $O^\thicksim(LM(q)+L\log p)$ arithmetic operations in $\RB$ plus  similar bit operations, where $LM(q)$ is the complexity of multiplying
two univariate polynomials with degrees $<q$.
By \cite{c}, we may assume $M(q)$ is $O(q\log q\log\log q)$. So it costs $O^\thicksim(Lq+L\log p)$ ring operations and similarly many bit operations.
The definition of $\S_{f_{(D,p,k)}}$ is the same as $\S_{f_{(D,p)}}$. The only difference is that when $j=k$, then replace $a_k$ by $a^{\modp(D^{k-1},p)+p}$.
\qed

\begin{remark}\label{rem-pr1}
From the above Lemma, although $\S_{f_{(D,p)}},\S_{f_{(D,p,k)}}$ are not   SLP procedures, we still can probe $f_{(D,p)}\ \modp\ (x^q-1)$ from them. Since in the following algorithms,  $p$ is $O^\thicksim(nT\log D)$ and $q$ is $O^\thicksim(T\log (nD))$, the complexity of the probing is $O^\thicksim(Lq)$. So we can still regard $\S_{f_{(D,p)}},\S_{f_{(D,p,k)}}$ as SLP procedures of length $L$.
\end{remark}

\vspace{11pt}
Let  \begin{eqnarray}\label{eq-t6}
 &&f_{(D,p)}^{\modp}=a_1x^{d_1}+a_2x^{d_2}+\cdots+a_r x^{d_r}\, (d_1< d_2<\cdots <d_r)\label{eq-mfdp}
\end{eqnarray}
Since $f_{(D,p)}\ \modp\ (x^p-1) = f_{(D,p,k)}\ \modp\ (x^p-1) =  f_{(D,p)}^{\modp}$, for $k=1,2,\dots,n$,   we can write
  \begin{eqnarray}\label{eq-t7}
 &&f_{(D,p)}=f_1+f_2+\cdots+f_r+g\label{eq-mfp}\\
 &&f_{(D,p,k)}=f_{k,1}+f_{k,2}+\cdots+f_{k,r}+g_k\nonumber
\end{eqnarray}
where
$f_i\ \modp\ (x^p-1)=f_{k,i}\ \modp\ (x^p-1)=a_ix^{d_i}$, $g\ \modp \ (x^p-1)=g_k\ \modp \ (x^p-1)=0$.
Similar to \bref{eq-uf}, we define the following key notation
\begin{eqnarray}
&&M^f_{D,p} =\{a_i x_1^{e_{i,1}}\cdots x_n^{e_{i,n}}| a_i \hbox{ is from } \bref{eq-mfdp} \hbox{  for some } i\in[1,r], \hbox{ and }  \cr
 &&\quad \hbox{M1}: f_i=a_ix^{u_i},f_{k,i}=a_ix^{b_{k,i}},k=1,2,\dots,n.\label{eq-uf2}\\ 
 &&\quad \hbox{M2}: e_{i,k}=\frac{b_{k,i}-u_i}{p}\in\N,k= 1,2,\dots,n.\cr
 &&\quad \hbox{M3}: u_i =e_{i,1}+e_{i,2}\modp(D,p)+\cdots+e_{i,n}\modp(D^{n-1},p)
 .\cr
 &&\quad \hbox{M4}:  \sum_{j=1}^n e_{i,j}< D. \}\nonumber
\end{eqnarray}

%

\begin{lemma}\label{lm-5}
Let $f=\sum_{i=1}^tc_im_i\in\RB[\X],D>\deg f$. If $c_im_i$ is not a collision in $f_{(D,p)}^ {\modp}$, then $c_im_i\in M^f_{D,p}$.
\end{lemma}
\proof
It suffices to show that $c_im_i$ satisfies the conditions of the definition of $M^f_{D,p}$. Assume $m_i=x_1^{e_1}x_2^{e_2}\cdots x_n^{e_n}$.
Since $c_im_i$ is not a collision in $f_{(D,p)}^{\modp}$,
without loss of generality, assume $(c_im_i)^{\modp}_{(D,p)} = a_1x^{d_1}$
and $d_1 = \modp(\sum_{j=1}^n e_jD^{j-1},p)$, where $a_1x^{d_1}$ is defined in \bref{eq-mfdp}.
It is easy to see that
$c_im_i$ is also not a collision in $f_{(D,p)}$ and in $f_{(D,p,k)}$.
Hence, $f_{1} = a_1 x^{u_{1}}$ for $u_{1}=\sum_{i=1}^n e_i\modp(D^{i-1},p)$;
$b_{k,1} = u_1 + pe_k$.
Clearly, M1, M2 and M3 are correct. Since $\deg (m_i)=\sum_{j=1}^n e_{i,j}< D$, M4 is correct.
\qed

Now we give the following algorithm to compute $M^f_{D,p}$, whose correctness is obvious.
\begin{alg}[MTerms]\label{alg-3}
\end{alg}

{\noindent\bf Input:}
Univariate polynomials $f^{\modp}_{(D,p)},f_{(D,p)},f_{(D,p,k)}$, where $k=1,2,\dots,n$, a prime $p$,   $D> \deg (f)$.

{\noindent\bf Output:} $M^f_{D,p}$.

\begin{description}
\item[Step 1:]
Write $f_{(D,p)}^{\modp}$, $f_{(D,p)}$, and $f_{(D,p,k)}$ as forms \bref{eq-t6} and \bref{eq-t7}
%
\begin{eqnarray*}
&&f_{(D,p)}^{\modp}=a_1x^{d_1}+a_2x^{d_2}+\cdots+a_rx^{d_r}\cr
&&f_{(D,p)}=f_1+f_2+\cdots+f_r+g\cr
&&f_{(D,p,k)}=f_{k,1}+f_{k,2}+\cdots+f_{k,r}+g_k, (k=1,2,\dots,n).
\end{eqnarray*}
%
%
\item[Step 2:] Let $S=\{\}$.
 For $i=1,2,\dots,r$ do
\begin{description}
\item[a:]
$\mathbf{If}$ one of $f_i,f_{1,i},\dots,f_{n,i}$ is not of the following form:
$f_i=a_ix^{u_i},f_{k,i}=a_ix^{b_{k,i}}$,
$\mathbf{then}$ break.

\item[b:]
Let $e_{i,k}=\frac{b_{k,i}-u_i}{p}$ for $k=1,2,\dots,n$.
 If $e_{i,k}\notin \N$, then break.

    \item[c:] If $u_i\neq e_{i,1}+e_{i,2}\modp(D,p)+\cdots+e_{i,n}\modp(D^{n-1},p)$, then break;

    \item[d:] If $\sum_{j=1}^ne_{i,j}\geq D$, then break;

    \item[e:]  Let $S=S\bigcup\{a_ix_1^{e_{i,1}}\cdots x_n^{e_{i,n}}\}$.
    \end{description}

\item[Step 3] Return $S$.
\end{description}

\begin{lemma}\label{the-3}
Algorithm \ref{alg-3} needs $O(nT)$ arithmetic operations in $\RB$ and $O^\thicksim(nT\log (pD))$ bit operations.
\end{lemma}
\proof
In Step 1, we need to write $f_{(D,p)}$ and $f_{(D,p,k)}$ as the desired form.
This can be done in three steps.
First, we perform the modular operation $\modp\ \ p$ on every degree of $f_{(D,p)},f_{(D,p,k)}$, which
costs $O(nT\log (pD))$ bit operations,
since each of  $f_{(D,p)},f_{(D,p,k)}$ has no more than $T$ terms
and the height of the degree is $O(\log (pD))$.
Second, we sort the terms of $f_{(D,p)}^{\modp}, f_{(D,p)},f_{(D,p,k)}$ into ascending order according to the new degree module $p$,
which costs $O^\thicksim(nT\log (p))$ bit operations, since the degrees are $< p$.
In order to check whether $f_i\ \modp\ (x^p-1)=f_{k,i}\ \modp\ (x^p-1)=a_ix^{d_i}$,
we need $O(Tn)$ operations over $\RB$.
Finally, $f_i, f_{k,i}$ can be obtained with $O(Tn)$ comparisons
of the degrees, which costs $O^\thicksim(nT\log (p))$ bit operations
and $O(Tn)$ $\RB$-operations.
So, the total complexity of Step 1 is $O^\thicksim(nT\log (pD))$ bit operations.

For Step 2, we first consider the complexity of one loop.
Since the height of the degrees of $f_{(D,p)}^{\modp}, f_{(D,p)},f_{(D,p,k)}$  are $O(\log (pD))$, Steps $\mathbf{a}$, $\mathbf{b}$, $\mathbf{c}$, and $\mathbf{d}$
costs $O(n\log (pD))$ bit operations.
Since we have at most $T$ loops, the total complexity is is $O(nT\log (pD))$  bit operations.

In $\mathbf{a}$ of Step 2, since $\#f_{(D,p)},\#f_{(D,p,k)}\leq T$, it totally needs $O^\thicksim(nT)$ bit operations to determine whether $\#f_i,\#(f_{k,i})=1$. To compare the coefficients of $f_i,f_{k,i}$, it needs $O(nT)$ arithmetic operations in $\RB$.
We prove the lemma
\qed

%
%
%
%
%
%
%
%
%
%

\subsection{The interpolation algorithm}
In this section, we give the interpolation algorithm for multivariate polynomial.
We first give a sub-algorithm, which computes $f_{(D,p)}$, and $f_{(D,p,k)}$ efficiently.

\begin{alg}[Substitution]\label{alg-mulp2}
\end{alg}

{\noindent\bf Input:}
A polynomial  $f\in\RB[\X]$, a prime $p$, a number $D\in N$ with $\deg f < D$.

{\noindent\bf Output:} The univariate polynomials $f_{(D,p)}$, and $f_{(D,p,k)},k=1,2,\dots,n$.

\begin{description}
\item[Step 1:] Assume $f=c_1m_1+c_2m_2+\cdots+c_tm_t$, where $m_i=x_1^{e_{i,1}}x_2^{e_{i,2}}\cdots x_n^{e_{i,n}},i=1,2,\dots,t$.

\item[Step 2:] Let $u_1=1$;

For $i=2,3,\dots,n$ do
$u_i=\modp(u_{i-1}D,p)$.

\item[Step 3:] Let $h_0=0$.
For $i=1,2,\dots,n$,
let $h_i=0$;

\item[Step 4:]
For $i=1,2,\dots,t$ do

\begin{description}
\item[a:] Let $d=0$.
\item[b:] For $k=1,2,\dots, n$,
let $d=d+e_{i,k}u_k$.
\item[c:] $h_0=h_0+c_ix^d$;

\item[d:] For $k=1,2,\dots,n$, let
 $h_k:=h_k+c_ix^{d+e_{i,k}p}$.
\end{description}

\item[Step 5:]
Return $h_0,h_i,i=1,2,\dots,n$;

\end{description}

\begin{lemma}\label{lm-m1}
Algorithm \ref{alg-mulp2} is correct. The complexity is $O^\thicksim(nt\log p+nt\log(D))$ bit operations and $O(nt)$ arithmetic operations in $\RB$.
\end{lemma}
\proof
In Step 2, $u_i=\modp(D^{i-1},p)$. In $\mathbf{b}$ of Step 4, $d$ is the degree of $m_{i(D,p)}$, so $h_0$ is $f_{(D,p)}$ after finishing Step 4. In $\mathbf{d}$, since $\deg (m_{i(D,p,k)})=\deg (m_{i(D,p)})+pe_{i,k}$, $h_k$ is $f_{(D,p,k)}$ after finishing Step 4. So the correctness is proved.

Now we analyse the complexity.
In Step 2, it needs $O(n\max\{\log D,\log p\})$ bit operations.
In $\mathbf{b}$ of Step 4, it needs $O(nt)$ arithmetic operations in $\Z$. Since $\deg (m_{i(D,p)})$ is $O(p\cdot\deg f)\le O(pD)$, the bit operations is $O(nt\log(pD))$.
In $\mathbf{c}$ and $\mathbf{d}$, it needs $O^\thicksim(nt\log (pD))$ bit operations and $O(nt)$ arithmetic operations in $\RB$.\qed

We now give the interpolation algorithm.
\begin{alg}[MPolySI]\label{alg-mi2}
\end{alg}

{\noindent\bf Input:} An  SLP  $\S_f$ that computes $f\in\RB[\X]$, $T\geq \# f$, $D>\deg f$.

{\noindent\bf Output:} The exact form of $f$.

\begin{description}
\item[Step 1:] Let $N_1=\max\{1,\lceil n(T-1)\log_2 D\rceil\},N_2=\lceil nT\log_2 D\rceil,N=\max\{4N_1,N_1+N_2-1\}$.

\item[Step 2:] Find the first $N$ different primes $p_1,p_2,\dots,p_{N}$.

\item[Step 3:]
 For $j=1,2,\dots,N$, probes
$f^{\modp}_{(D,p_j)}$. Let $f^{\modp}_j=f^{\modp}_{(D,p_j)}$.

\item[Step 4:] 
    Let $h=0$ and $T_1=T$.

\item[Step 5:] Loop

\begin{description}

\item[5.1:] Let $\alpha=\max\{\#f^{\modp}_j|j=1,2,\dots,N\}$ and $j_0$ the smallest number such that $\#f^{\modp}_{j_0}=\alpha$.

\item[5.2:] If $\alpha=0$ return $h$.

\item[5.3:] $\{f^*,f^*_1,\dots,f^*_n\}=\mathbf{Substitution}(h,p_{j_0},D)$.

\item[5.4:]
Let $f_{j_0}=\mathbf{UIPoly}(\S_{f_{(D,p_{j_0})}},f^*,T,T_1,Dp_{j_0})$.

\item[5.5:] For $k=1,2,\dots,n$, let $g_k=\mathbf{UIPoly}(\S_{f_{(D,p_{j_0},k)}},f^*_k,T,T_1,2Dp_{j_0})$.

\item[5.6:] Let $M^{f-h}_{D,p_{j_0}}:=\mathbf{MTerms}(f^{\modp}_{j_0}, f_{j_0},g_1,g_2,\dots,g_n,p_{j_0},D)$.

\item[5.7:] Let $s=0$. For each $u\in M^{f-h}_{D,p_{j_0}}$,
if
$$\#\{j\,|\, \#(f^{\modp}_j-u^{\modp}_{(D,p_j)})<\#(f^{\modp}_j),j=1,\dots,N_1+N_2-1 \}\ge N_2,$$
then $s:=s + u$.

\item[5.8:] Let $h=h+s$, $T_1=T_1-\#s$, $N_1=\max\{1,\lceil n(T_1-1)\log_2 D\rceil\},N_2=\lceil nT_1\log_2 D\rceil,N=\max\{4N_1,N_1+N_2-1\}$.

\item[5.9:] For $j=1,2,\dots,N$, let $f^{\modp}_j=f^{\modp}_j-s^{\modp}_{(D,p_{j})}$.

%
\end{description}


\end{description}

\begin{theorem}\label{the-2}
Algorithm \ref{alg-mi2}   finds $f$ using   $O^\thicksim(Ln^2T^2\log^2 D+LnT\log^3 D)$ ring operations in $\RB$ and similar bit operations.
\end{theorem}
\proof
The algorithm is quite similar to the univariate interpolation Algorithm \ref{alg-2}.
So, we will only give the sketch of the proof and give detailed proof
only for those steps and are essentially different from that in Algorithm \ref{alg-2}.
By Theorem \ref{lm-21} and Lemma \ref{lm-5}, at least half of terms of $f-h$ are in $M^{f-h}_{D,p_{j_0}}$ obtained in Step 5.6.
In  Step 5.7, Theorem \ref{lm-2} is used to select the elements of $M_{f-h}$ from $M^{f-h}_{D,p_{j_0}}$.
So at least half of the terms of $f-h$ will be found in each loop of Step 5. Then the correctness of the algorithm is proved.

We now analyse the complexity of the algorithm, which comes from that of Steps 3 and 5.
In Step 2, since the bit complexity of finding the first $N$ primes is $O(N\log^2N\log\log N)$ by \cite[p.500,Thm.18.10]{12} and $N$ is $O^\thicksim(nT\log D)$, the bit complexity of Step 2 is $O^\thicksim(nT\log D)$.

In Step 3, we  probe  $\S_f$ for $O(nT\log D)$ times. Since $p_i$ is  $O^\thicksim(nT\log D)$, the cost of probes is $O^\thicksim(Ln^2T^2\log^2D)$ ring and bit operations.

We now consider the complexity of one loop for Step 5.
In Step 5.1, since $N$ is of $O(nT\log D)$ and  $\#f^{\modp}_{j_0}\leq T$, the bit complexity is $O^\thicksim(nT^2\log D)$.

In Steps 5.3,
by Lemma \ref{lm-m1},
since $p_i$ is  $O^\thicksim(nT\log D)$,
the complexity is $O^\thicksim(n\log D+nT\log p_{j_0}+nT\log D)=O^\thicksim(nT\log D)$ bit operations and $O(nT)$ arithmetic operations in $\RB$.

In Steps 5.4 and 5.5, by Remark \ref{lm-pr1}, we can regard
$\S_{f_{(D,p_{j_0})}}$ and $\S_{f_{(D,p_{j_0},k)}}$ as SLP procedures of length $L$.
Since the numbers of terms and degrees of $(f-h)_{(D,p_{j_0})}$, $(f-h)_{(D,p_{j_0},k)}$ are bounded by $O(T)$ and $O^\thicksim(nTD)$, by Theorem \ref{the-1}, the complexity is $O^\thicksim(LnT^2\log^2 D+LnT\log^3 D)$ ring and bit operations.

In Step 5.6, by Theorem \ref{the-3}, the complexity is $O^\thicksim(nT\log D)$ bit operations and $O(nT)$ ring operations.

In Step 5.7, to all the $u^{\modp}_{(D,p_j)}$, we need $O^\thicksim(n(N_1+N_2)\#M^{f-h}_{D,p_{j_0}}\log (Dp)$ bit operations.
The proof for rest of this step is similar to that of  Step 5.5 of Algorithm \ref{alg-2}.
The complexity is $O^\thicksim(n(N_1+N_2)\#M^{f-h}_{D,p_{j_0}}\log (Dp_j)+ \#M^{f-h}_{D,p_{j_0}}\log T$ $(N_1+N_2)\log (Dp_j))$ bit operations and $O(\#M^{f-h}_{D,p_{j_0}} (N_1+N_2))$ ring operations. Since $\#M^{f-h}_{D,p_{j_0}}\leq T$, $p_j=O^\thicksim(nT\log D)$, and  $N_1+N_2 = O(nT\log D)$, it needs  $O^\thicksim(n^2T^2\log^2 D)$ bit operations and $O^\thicksim(nT^2\log D)$ ring operations.

In Step 5.9, we need $n\# s$ operations in $\Z$ to obtain $s^{\modp}_{(D,p_j)}$. Subtract $s^{\modp}_{(D,p_j)}$ from $f_j$ needs $\# s\log T$ operations in $\Z$ and $\#s$ arithmetic operation in $\RB$. Since the height of the data is $O(\log (nTD))$ and we need update $N$ polynomials,  the complexity is $O((n\#s\log (nTD)+\#s\log T$ $\log (nTD))N)$ bit operations and $O(\#sN)$ ring operations.
Since the sum of $\#s$ is $t$, it total costs $O^\thicksim(nT^2\log D)$ ring operations and $O^\thicksim(n^2T^2\log^2 D)$ bit operations.

%
Then the total complexity of one loop of Step 5 is
 $O^\thicksim(LnT^2\log^2 D+LnT\log^3 D)$ ring  operations  
 and
$O^\thicksim(LnT^2\log^2 D+LnT\log^3 D+n^2T^2\log^2 D)$ bit operations,
which come from Steps 5.5, 5.7, and 5.9.
Since every loop of Step 5 finds at least half of the terms in $f-h$,
the loop runs at most $O(\log T)$ times.
So, the total complexity of Step 5 is
$O^\thicksim(LnT^2\log^2 D+LnT\log^3 D)$ ring  operations
 and
$O^\thicksim(LnT^2\log^2 D+LnT\log^3 D+n^2T^2\log^2 D)$ bit operations.
Plus the complexity of Step 3, the complexity of the algorithm is
$O^\thicksim(Ln^2T^2\log^2 D+LnT\log^3 D)$ ring   and   bit operations,
  which are from Step 3 and Step 5.5.
\qed

\section{Monte Carlo algorithm for multivariate polynomials}
In this section, we give Monte Carlo interpolation algorithms for multivariate polynomials, which could be considered as probabilistic versions of Algorithm \ref{alg-mi2}.

The following theorem shows how to use a probabilistic method
to obtain a $p$ such that the number of collision terms of $f$  in $f^{\modp}_{(D,p)}$ is very small, which is a probabilistic version of Theorem \ref{lm-21}. 
\begin{theorem}\label{them-1}
Let $f=\sum_{i=1}^tc_im_i\in\RB[\X]$, $T\geq \# f,D>\deg f$, $N=\max\{1,\lfloor \frac{32}{5}n(T-1)\log_2 D\rfloor\}$, and $p_1,p_2,\dots,p_{2N}$ be $2N$ different primes, $0<\varepsilon<1$, $k=\lceil\log_2(\frac{1}{\varepsilon})\rceil$. If $j_1,j_2,\dots,j_k$ are randomly chosen from $[1,2N]$  and $j_0\in\{j_1,\dots,j_k\}$ is the integer such that $\#f^{\modp}_{(D,p_{j_0})}=\max\{\#f^{\modp}_{(D,p_{j_1})},\#f^{\modp}_{(D,p_{j_2})},\\ \dots,\#f^{\modp}_{(D,p_{j_k})}\}$. Then with probability $\ge  1-\varepsilon$, $\mathcal{C}^f_{(D,p_{j_0})}\leq \lfloor\frac{5}{16}T\rfloor$.
\end{theorem}
\proof
If $T=1$, then $N_1=1$, the proof is obvious. So we assume $T\geq 2$, then $N_1=\lfloor \frac{32}{5}n(T-1)\log_2 D\rfloor$.
First we claim that if randomly choose an integer $j$ in $[1,2N]$, then with probability at least $\frac12$, $\mathcal{C}^f_{(D,p_j)}< \frac{5}{32}T$.
It suffices  to show that there exist at least $N$ integers $j$ in $[1,2N]$ such that $\mathcal{C}^f_{(D,p_j)}<\frac{5}{32}T$.
We prove this by contradiction. Assume $\alpha_1,\alpha_2,\dots,\alpha_{N+1}$ are $N+1$ different integers in $[1,2N]$ such that $\mathcal{C}^f_{(D,p_{\alpha_i})}\geq \frac{5}{32}T$.
  By Lemma \ref{lm-6}, $p_{\alpha_i}^{\lceil \mathcal{C}^f_{(D,p_{\alpha_i})}/2\rceil}$ divides $A$, where $A$ is defined in Lemma \ref{lm-6}.
   Since $p_{\alpha_i}$ are different and $\mathcal{C}^f_{(D,p_{\alpha_i})}\geq \frac{5}{32}T$,  $(p_{\alpha_1}p_{\alpha_2}\cdots p_{\alpha_{N+1}})^{\lceil\frac{5}{64}T\rceil}$ divides $A$. Now we have $(p_{\alpha_1}p_{\alpha_2}\cdots p_{\alpha_{N+1}})^{\lceil\frac{5}{64}T\rceil}\geq 2^{(N+1)\lceil\frac{5}{64}T\rceil}\geq 2^{\lceil\frac{32}{5}n(T-1)\log_2 D\rceil\lceil\frac{5}{64}T\rceil}\geq 2^{\frac12 nT(T-1)\log_2 D}=D^{\frac{1}{2}nT(T-1)}$,
    which contradicts to $A\leq (D^n-1)^{\frac{1}{2}t(t-1)}$. So in $\{p_1,p_2,\cdots,p_{2N}\}$, there are at least $N$ primes $p_i$ such that $\mathcal{C}^f_{(D,p_i)}<\frac{5}{32}T$. We have proved the   claim.

If there exists at least one $j_i$ in $\{j_1,j_2,\dots,j_k\}$ such that $\mathcal{C}^f_{(D,p_{j_i})}<\frac{5}{32}T$, then by Corollary \ref{cor-1}, $\mathcal{C}^f_{(D,p_{j_0})}\leq 2\mathcal{C}^f_{(D,p_{j_i})}<\frac{5}{16}T$.
So $\mathcal{C}^f_{(D,p_{j_0})}\geq \frac{5}{16}T$ only when all $\mathcal{C}^f_{(D,p_{j_i})}\geq \frac{5}{32}T,i=1,2,\dots,k$, 
by the claim just proved,
the probability for this to happen is at most $(\frac{1}{2})^k\leq (\frac12)^{\log_2 \frac{1}{\varepsilon}}=\varepsilon$.
Since $\mathcal{C}^f_{(D,p_{j_0})}<\frac{5}{16}T$ implies  that $\mathcal{C}^f_{(D,p_{j_0})}\leq \lfloor \frac{5}{16}T\rfloor$, the probability of $\mathcal{C}^f_{(D,p_{j_0})}\leq \lfloor\frac{5}{16}T\rfloor$ is at least $1-\varepsilon$. The theorem is proved.
\qed

\begin{remark}
Note that the result $\mathcal{C}^f_{(D,p_{j_0})}\leq \lfloor\frac{5}{16}T\rfloor$ of Theorem \ref{them-1} is different
with that of Theorem \ref{lm-21}.
To find a $p$ such that
at least $\lceil\frac{t}{2}\rceil$ of the terms of $f$ are not collisions in $f^{\modp}_{(D,p_{j_0})}$ is not enough for our probabilistic algorithm.
\end{remark}

\begin{lemma}\label{them-4}
If $\varepsilon\in(0,1)$ and $a\geq 1$, then  $(1-\varepsilon)^a\geq 1-a\varepsilon$.
\end{lemma}
\proof
By Taylor expansion, we have $(1+x)^a=1+ax+a(a-1)\frac{(1+\theta x)^{a-2}}{2}$, where $\theta\in(0,1)$. Now we let $x=-\varepsilon$, then $(1-\varepsilon)^{a}=1-a\varepsilon+a(a-1)\frac{(1-\theta \varepsilon)^{a-2}}{2}$.   Since $\theta\in(0,1)$, $\varepsilon\in(0,1)$ and $a\geq 1$, we have $a(a-1)\frac{(1-\theta \varepsilon)^{a-2}}{2}\geq 0$.
So we have $(1-\varepsilon)^{a}\geq1-a\varepsilon$. \qed

We first consider interpolation over an arbitrary computable ring.
For the univariate interpolation algorithm, we use the following algorithm given in \cite{2}, which is the fastest known probabilistic  algorithm over arbitrary rings.

\begin{theorem}\cite{2}\label{them-2}
Let $f\in\RB[x]$, where $\RB$ is any ring. Given any SLP of length $L$ that computes $f$, and bounds $T$ and $D$ for the sparsity and degree of $f$, one can find all coefficients and exponents of $f$ using $O^\thicksim(LT\log^3 D+LT\log D\log\frac{1}{\nu})$ ring operations in $\RB$, plus a similar number of bit operations. The algorithm is probabilistic of the Monte Carlo type: it can generate random bits at unit cost and on any invocation returns the correct answer with probability greater than $1-\nu$, for a user-supplied tolerance $0<\nu<1$.
\end{theorem}

We use  $\mathbf{PUniPoly}_1(\S_f,f^*,T_1,D,\nu)$ to denote the algorithm in Theorem \ref{them-2}, where $f^*$ is a current approximation to $f$ and $\#(f-f^*)\leq T_1,\#f^*\leq T, \max\{\deg f^*,\deg f\}<D$.

Now we give an algorithm which interpolates at least half of the terms.

\begin{alg}[HalfPoly]\label{alg-4}
\end{alg}

{\noindent\bf Input:} An SLP $\S_f$ that computes $f$, $f^*\in\RB[\X]$, $T\geq \max\{\# f,\#f^*\}$, $T_1\geq
\#(f-f^*)$, $D>\max\{\deg f,\deg f^*\}$, a tolerance $\nu$ such that $0<\nu<1$.

{\noindent\bf Output:} With probability $\geq1-\nu$, return a polynomial $f^{**}$ such that $\#(f-f^*-f^{**})\leq \lfloor\frac{T_1}{2}\rfloor$.

\begin{description}
\item[Step 1:] Let $N=\max\{1,\lceil \frac{32}{5}n(T_1-1)\log_2 D\rceil\}$, $\varepsilon=\frac{\nu}{n+1}$, and 
    $k=\lceil\log_2 \frac{1}{\varepsilon}\rceil$. Find the first $2N$ primes $p_1,p_2,\dots,p_{2N}$.

\item[Step 2:] Let $j_1,j_2,\dots,j_k$ be randomly chosen from $[1,2N]$. Delete the repeated numbers, we still denote these integers as $j_1,j_2,\dots,j_k$.

\item[Step 3:] For $i=1,2,\dots,k$, probe $f^{\modp}_{(D,p_{j_i})}$. Let $g_i=f^{\modp}_{(D,p_{j_i})}-f^{*\modp}_{(D,p_{j_i})}$.

\item[Step 4:] Let $\alpha=\max\{\#g_i|i=1,2,\dots,k\}$ and $j_0$ satisfying $\#g_{j_0}=\alpha$. If $\alpha\geq T$, then return failure.

\item[Step 5:] Let $\{q,q_1,\dots,q_n\}=\mathbf{Substitution}(f^*,p_{j_0},D)$.

\item[Step 6:]
Let
$\eta=\mathbf{PUniPoly}_1(\S_{f_{(D,p_{j_0})}},q,T_1,p_{j_0}D,\varepsilon)$,
$\eta_i=\mathbf{PUniPoly}_1(\S_{f_{(D,p_{j_0},i)}},q_i,T_1,2p_{j_0}D,\varepsilon),$ $i=1,\dots,n$.
If $\eta$ or $\eta_i$ is failure, then return failure.

\item[Step 7:]
Let $M=\mathbf{MTerms}(g_{j_0},\eta,\eta_1,\eta_2,\dots,\eta_n,D,p_{j_0})$.

\item[Step 8:]
Return $f^{**}=\sum_{s\in M} s$.
\end{description}

\begin{lemma}\label{them-3}
Algorithm \ref{alg-4} computes $f^{**}$ such that $\#(f-f^*-f^{**})\leq \lfloor\frac{T_1}{2}\rfloor$ with probability at least $1-\nu$.
The algorithm costs
$O^\thicksim(LnT\log^3D+LnT\log D\log \frac{1}{\nu})$ ring operations in $\RB$ and a similar many bit operations.
\end{lemma}
\proof
We first show that Algorithm \ref{alg-4} returns the polynomial $f^{**}$ such that $\#(f-f^*-f^{**})\leq \lfloor\frac{T_1}{2}\rfloor$ with probability $1-\nu$.
In Step 4, by Theorem \ref{them-1}, with probability $1-\varepsilon$, $\mathcal{C}^{f-f^*}_{(D,p_{j_0})}\leq \lfloor\frac{5}{16}T_1\rfloor$.
If $j_0$ satisfies $\mathcal{C}^{f-f^*}_{(D,p_{j_0})}\leq \lfloor\frac{5}{16}T_1\rfloor$ and $\eta=(f-f^*)_{(D,p_{j_0})},\eta_i=(f-f^*)_{(D,p_{j_0},i)},i=1,2,\dots,n$, then by Lemma \ref{lm-5},  $f-f^*$ contains at most  $\lfloor\frac{5}{16}T_1\rfloor$ terms which are not in $f^{**}$.
Since the terms of $f^{**}$ which are not in $M_{f-f^*}$ come from at least two terms in $f-f^*$,  there exist at most $\frac12\lfloor\frac{5}{16}T_1\rfloor$ terms of $f^{**}$ not in  $f-f^*$. So $\#(f-f^*-f^{**})\leq \lfloor\frac{5}{16}T_1\rfloor+\frac12\lfloor\frac{5}{16}T_1\rfloor\leq \frac{15}{32}T_1< \frac12T_1$
and  we have $\#(f-f^*-f^{**})\leq \lfloor \frac12T_1\rfloor$.
In Step6,  by Theorem \ref{them-2}, the probability of obtaining the correct
$(f-f^*)_{(D,p_{j_0})}$ $((f-f^*)_{(D,p_{j_0},i)})$, or $\eta=(f-f^*)_{(D,p_{j_0})}$ $(\eta_i=(f-f^*)_{(D,p_{j_0},i)})$, is $\geq1-\varepsilon$, so the probability of obtaining correct polynomials $(f-f^*)_{(D,p_{j_0})}$
 and $((f-f^*)_{(D,p_{j_0},i)}), i=1,\ldots,n$
 is $\geq(1-\varepsilon)^{n+2}$. By Lemma \ref{them-4}, $(1-\varepsilon)^{n+1}\geq1-(n+1)\varepsilon$. Since $\varepsilon=\frac{\nu}{n+1}$, $(1-\varepsilon)^{n+1}\geq1-\nu$.
Hence, with  probability $\geq1-\nu$, we obtain an $f^{**}$
satisfying $\#(f-f^*-f^{**})\leq \lfloor\frac{T_1}{2}\rfloor$.
The correctness of the lemma is proved.

Now we analyse the complexity.
 Since the bit complexity of finding the first $2N$ primes is $O(N\log^2N\log\log N)$ by \cite[p.500,Thm.18.10]{12} and $N$ is $O^\thicksim(nT\log D)$, the bit complexity of Step 1 is $O^\thicksim(nT\log D)$.

 In Step 2, since probabilistic machines flip coins to decide binary digits, each of these random choices can be simulated with a machine with complexity $O(\log 2N)$. So the complexity of Step 2 is $O(\log \frac{1}{\varepsilon}\log (nT\log D))$ bit operations.
 Since $O(\log\frac 1\varepsilon)$ is $O(\log n+\log \frac{1}{\nu})$,  the bit complexity of Step 2 is $O(\log n\log (nT\log D)+\log (nT\log D)\log \frac{1}{\nu})$.

In Step 3, we probe $k=\lceil\log_2 \frac{1}{\varepsilon}\rceil$ times for $f$. Since $p_{j_i}$ is of $O^\thicksim(nT\log D)$, the cost of the probes is $O^\thicksim(LnT\log D\log \frac{1}{\varepsilon})$ ring operations and a similar many bit operations. So the complexity of step 3 is  $O^\thicksim(LnT\log D\log \frac{1}{\nu})$ arithmetic operations in $\RB$ and a similar number of  bit operations.

In Step 4, we find the integer $j_0$. Since $\#(f-f^*)^{\modp}_{(D,p_{j_i})}\leq T$, it needs at most $O^\thicksim(T\log \frac{1}{\varepsilon})$ bit operations to compute all $\#(f-f^*)^{\modp}_{(D,p_{j_i})},i=1,2,\dots,k$.
Find $j_0$ needs $O^\thicksim(T)$ bit operations. So the bit complexity of Step 4 is $O^\thicksim(T\log n+T\log \frac{1}{\nu})$.

In Step 5, by Lemma \ref{lm-m1}, it needs $O^\thicksim(nT\log D)$ bit operations and $O(nT)$ ring operations.

In Step 6, we call $n+1$ times Algorithm $\mathbf{PUniPoly}_1$.
Since the terms and degrees of $(f-f^*)_{(D,p_{j_0})}$, $(f-f^*)_{(D,p_{j_i},k)}$ are respectively bounded by $T$ and $2p_{j_0}D$, by Theorem \ref{them-2}, the complexity of Step 6 is $O^\thicksim(LnT\log^3(p_{j_0}D)+LnT\log (p_{j_0}D)\log \frac{1}{\varepsilon}))$ arithmetic operations in $\RB$ and plus a similar number of bit operations.
Since $\varepsilon=\frac{\nu}{n+1}$, $p_{j_0}$ is $O^\thicksim(nT\log D)$, and $2p_{j_0}D$ is $O^\thicksim(nTD)$,  the complexity of step 6 is $O^\thicksim(LnT\log^3D+LnT\log D\log \frac{1}{\nu}))$ ring operations and similar bit operations.


In Step 7, by Theorem \ref{the-3}, the complexity is $O(nT)$ arithmetic operation in ring $\RB$ and $O^\thicksim(nT\log pD)$ bit operations.

It is easy to see that the complexity is dominated by  Step 3 and Step 6.
The theorem is proved.\qed

We now give the complete interpolation algorithm.
\begin{alg}[MCMulPoly]\label{alg-m3}
\end{alg}

{\noindent\bf Input:}
An SLP   $S_f$ that computes $f\in\RB[\X]$,  $T> \# f$, $D> \deg f$, and $\mu\in(0,1)$.

{\noindent\bf Output:} With probability at least $1-\mu$, return $f$.

\begin{description}
\item[Step 1:] Let $h=0,T_1=T,\nu=\frac{\mu}{\lceil\log_2 T\rceil+1}$.

\item[Step 2:] While $T_1>0$ do

\begin{description}
\item[b:] Let $g=\mathbf{HalfPoly}(\S_f,h,T,T_1,D,\nu)$. If $g=failure$, then return failure.

\item[c:] Let $h=h+g$, $T_1=\lfloor\frac{T_1}{2}\rfloor$.
\end{description}

\item[Step 3:] Return $h$.

\end{description}

\begin{theorem}\label{them-5}
Algorithm \ref{alg-m3} computes $f$ with probability $\ge 1-\mu$.
The algorithm costs
$O^\thicksim(LnT\\\log^3D+LnT\log D\log \frac{1}{\mu})$ ring operations in $\RB$ and a similar number of bit operations.
\end{theorem}
\proof
In Step 2, by Lemma \ref{them-3}, $\#(f-h-g)\leq \lfloor \frac{T_1}{2}\rfloor$ with probability $\ge 1-\nu$.
By Lemma \ref{them-4},  Step 2 will run at most $k= \lceil\log_2 T\rceil+1$ times and return the correct $f$ with  probability $\geq(1-\nu)^k\geq 1-\mu$.
The correctness of the theorem is proved.

 Now we analyse the complexity.
 In Step 2, we call at most $O(\log T)$ times Algorithm $\mathbf{HalfPoly}$. Since the terms and degrees of $f-h,f$ are respectively bounded by $T$ and $D$, by Theorem \ref{them-3}, the complexity of Step 2 is $O^\thicksim(LnT\log^3D+LnT\log D\log \frac{1}{\nu}))$ ring and bit operations.
 Since $\nu=\frac{\mu}{\lceil\log_2 T\rceil+1}$, the complexity of Step 2 is $O^\thicksim(LnT\log^3D+LnT\log D\log \frac{1}{\mu}))$ ring and bit operations.
The theorem is proved.\qed
\begin{remark}
Set $\mu=1/4$. Then
Algorithm \ref{alg-m3} computes $f$ with probability at lest $\frac34$.
The algorithm costs $O^\thicksim(LnT\log^3D)$ ring operations in $\RB$ and a similar number of bit operations.
\end{remark}

We now consider interpolation over finite fields.
%
In \cite{2i}, Arnold, Giesbrecht $\&$ Roche gave a new univariate interpolation algorithm  for the finite field with better complexities.
%
%
\begin{theorem}\cite{2i}\label{them-6}
Let $f\in\F_q[x]$ with at most $T$ non-zero terms and degree at most $D$, and let $0<\eta\leq\frac12$. Suppose we are given an SLP $\S_f$ of length $L$ that computes $f$. Then there exists an algorithm   that interpolates $f$, with probability at least $1-\eta$, with a cost of $O^\thicksim(LT\log^2 D(\log D+\log q)\log\frac{1}{\eta})$ bit operations.
\end{theorem}

We use $\mathbf{PUniPoly}_2(f,f^*,T_1,D,\eta)$ to denote the algorithm in Theorem \ref{them-6}, where $f^*$ is a current approximation to $f$ and $\#(f-f^*)\leq T_1,\#f^*\leq T, \deg f^*<D,\deg f<D$.

Replacing Algorithm $\mathbf{PUniPoly}_1$ with Algorithm $\mathbf{PUniPoly}_2$ in Step 6 of Algorithm $\mathbf{HalfPoly}$, we obtain a multivariate interpolation algorithm for finite fields. We assume $\S_f$ can evaluate in an extension field of $F_q$
and  have the following result.

\begin{theorem}\label{them-7}
Let $f\in\F_q[\X]$ be given as an SLP, with at most $T$ non-zero terms and degree at most $D$, and let $0\le \mu<1/2$.  Then we can interpolate $f$, with probability at least $1-\mu$, with a cost of $O^\thicksim(LnT\log^2 D(\log D+\log q)\log \frac{1}{\mu})$ bit operations.
\end{theorem}
\proof
The proof is the same as that of Theorem \ref{them-5}.
The only difference is to use Theorem \ref{them-7} instead of 
Theorem \ref{them-2}.
\qed

\section{Experimental results}

In this section, practical performances of the new algorithms
implemented in Maple will be presented.
The data are collected on a desktop with Windows system,
3.60GHz Core i7-4790 CPU, and 8GB RAM memory.
The Maple codes can be found in
\begin{verbatim}
http://www.mmrc.iss.ac.cn/~xgao/software/slppoly.zip
\end{verbatim}

We randomly construct five polynomials, then regard them as SLP polynomials
and reconstruct them with the Algorithms \ref{alg-2} , \ref{alg-mi2} and \ref{alg-m3}. We do not collect the time of probes. We just test the time of recovering $f$ from the univariate polynomial $f^{\modp}_{(D,p)},f_{(D,p)}$ and $f^{k}_{(D,p)}$. The average times for the five examples are collected.

\begin{figure}[ht]
\begin{minipage}[t]{0.31\linewidth}
\centering
\includegraphics[scale=0.22]{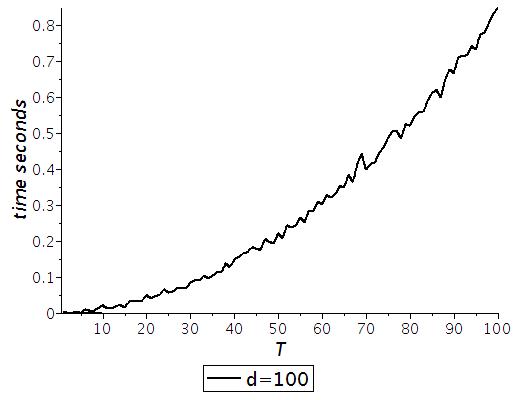}
\caption[UT.jpg]{\small Horizontal-axis: $T$} \label{fig1}
\end{minipage}\quad
\begin{minipage}[t]{0.31\linewidth}
\centering
\includegraphics[scale=0.21]{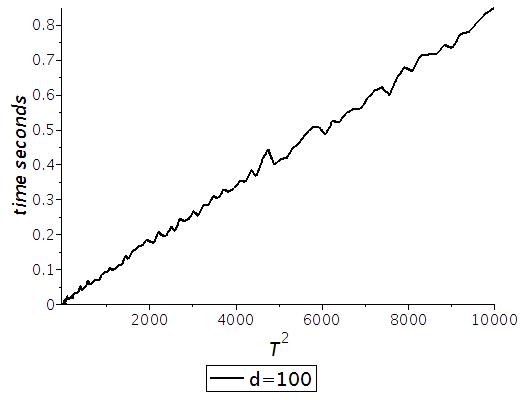}
\caption[UT2.jpg]{\small Horizontal-axis: $T^2$ }\label{fig2}
\end{minipage}\quad
\begin{minipage}[t]{0.31\linewidth}
\centering
\includegraphics[scale=0.22]{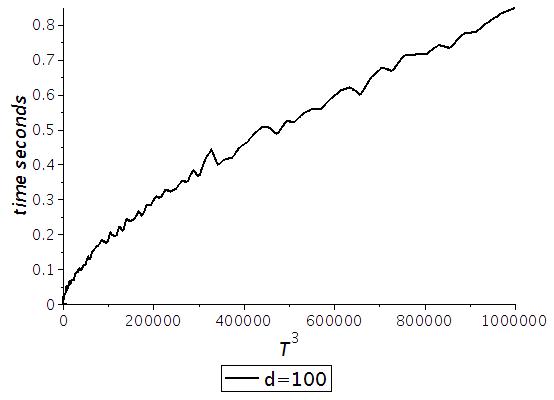}
\caption{\small Horizontal-axis: $T^3$} \label{fig3}
\end{minipage}\\
\begin{minipage}[t]{0.48\linewidth}
\centering
\includegraphics[scale=0.21]{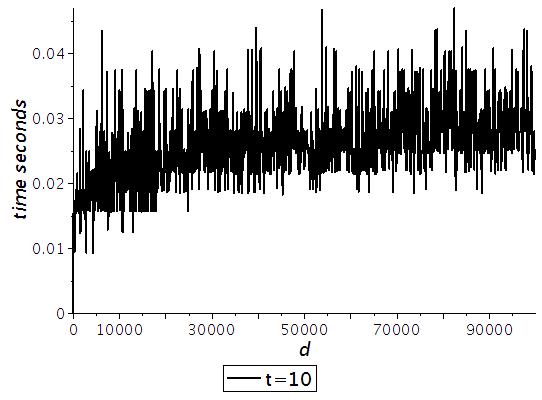}
\caption{Horizontal-axis: $d$ }\label{fig4}
\end{minipage}
\begin{minipage}[t]{0.48\linewidth}
\centering
\includegraphics[scale=0.21]{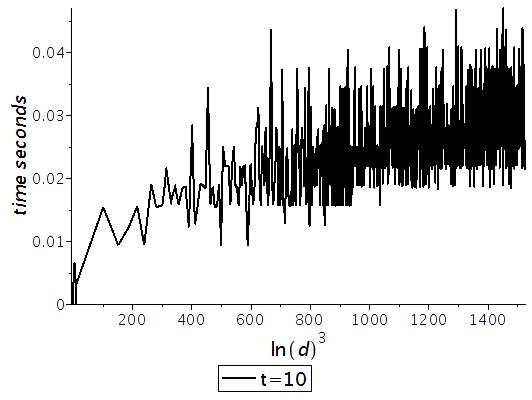}
\caption{Horizontal-axis: $\log^3 d$ }\label{fig5}
\end{minipage}
\end{figure}

For Algorithm \ref{alg-2}, the relations between the running times and   $T,T^2,T^3$ are respectively given in Figures \ref{fig1}, \ref{fig2}, and \ref{fig3}, where the parameter $d$ is fixed.
The relations between the running times and  $D,\log D$ are respectively given in Figures \ref{fig4} and \ref{fig5}, where the parameter $t$ is fixed.
Figures \ref{fig4} and \ref{fig5} show that the complexity of
Algorithm \ref{alg-2} is sensitive to $D$.
Overall, these figures are basically in accordance with
the theoretical complexity bound $O^\thicksim(LT^2\log^2 D+LT\log^3 D)$ for Algorithm \ref{alg-2}.

\begin{figure}[ht]
\begin{minipage}[t]{0.31\linewidth}
\centering
\includegraphics[scale=0.22]{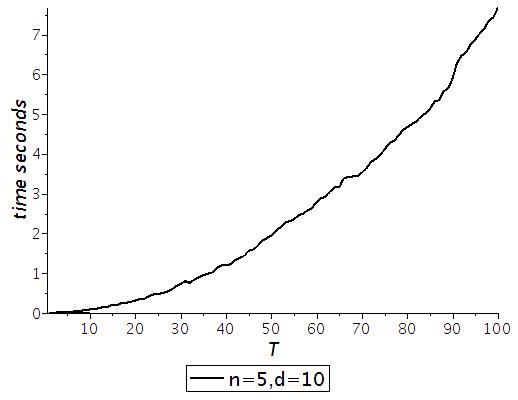}
\caption[UT.jpg]{\small Horizontal-axis: $T$} \label{fig6}
\end{minipage}\quad
\begin{minipage}[t]{0.31\linewidth}
\centering
\includegraphics[scale=0.21]{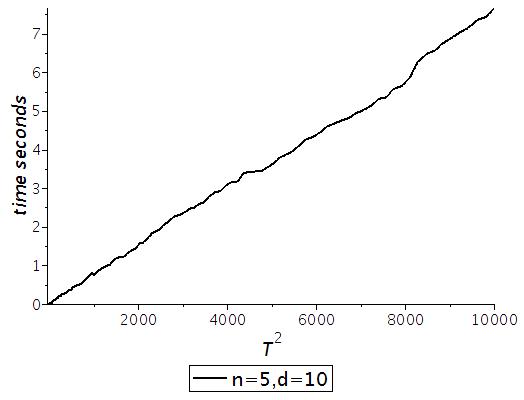}
\caption[UT2.jpg]{\small Horizontal-axis: $T^2$ }\label{fig7}
\end{minipage}\quad
\begin{minipage}[t]{0.31\linewidth}
\centering
\includegraphics[scale=0.22]{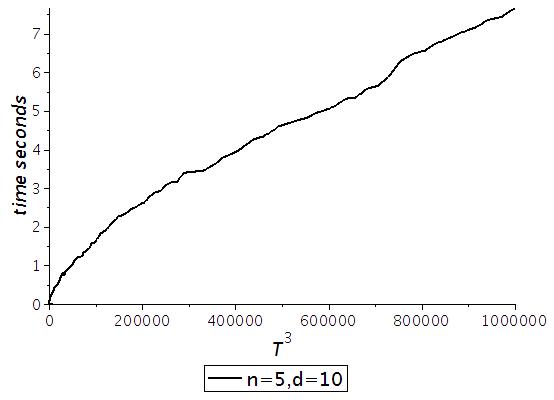}
\caption{\small Horizontal-axis: $T^3$} \label{fig8}
\end{minipage}\\
\end{figure}

\begin{figure}
\begin{minipage}[t]{0.48\linewidth}
\centering
\includegraphics[scale=0.21]{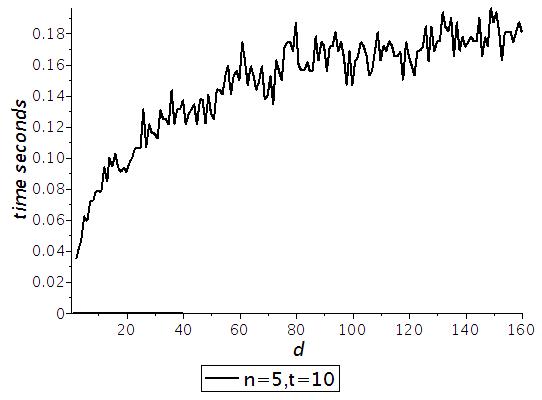}
\caption{Horizontal-axis: $d$ }\label{fig9}
\end{minipage}
\begin{minipage}[t]{0.48\linewidth}
\centering
\includegraphics[scale=0.21]{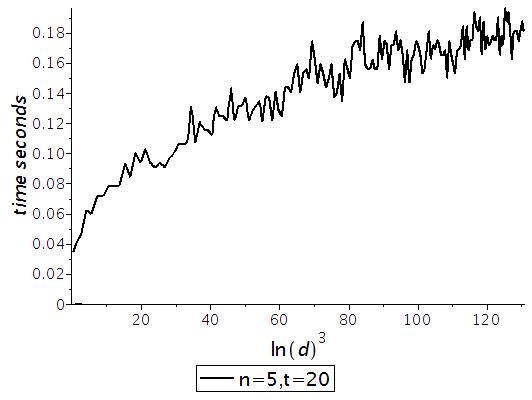}
\caption{Horizontal-axis: $\ln^3 d$ }\label{fig10}
\end{minipage}
\end{figure}

\begin{figure}
\begin{minipage}[t]{0.48\linewidth}
\centering
\includegraphics[scale=0.21]{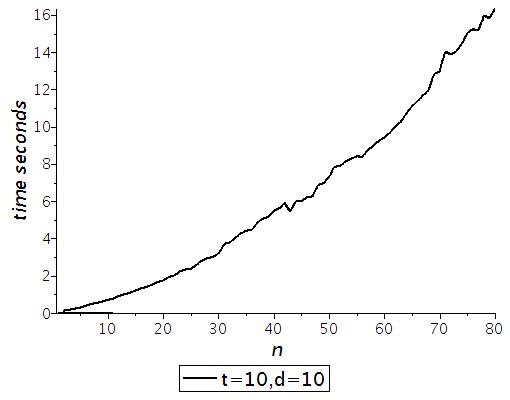}
\caption{Horizontal-axis: $n$ }\label{fig11}
\end{minipage}
\begin{minipage}[t]{0.48\linewidth}
\centering
\includegraphics[scale=0.21]{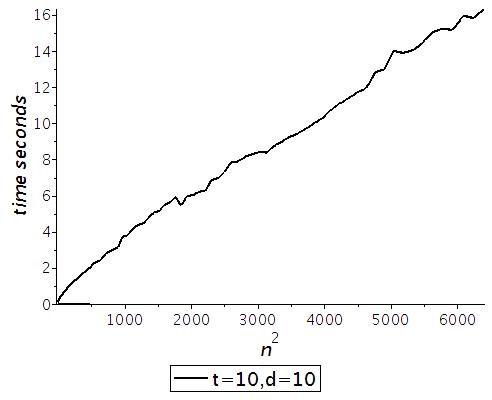}
\caption{Horizontal-axis: $n^2$ }\label{fig12}
\end{minipage}
\end{figure}

For Algorithm \ref{alg-mi2}, the relations between the running times  and $T,T^2,T^3$ are
respectively given in Figures \ref{fig6}, \ref{fig7}, \ref{fig8}, where the parameters $n,d$ are fixed.
Similarly, the relations between the running times and $d,\log d, n, n^2$
are respectively given in Figures Figures \ref{fig9}, \ref{fig10}, \ref{fig11}, \ref{fig12}.
From these, we can see that the practical performances 
are basically in accordance with
the theoretical complexity bound $O^\thicksim(Ln^2T^2\log^2 D+LnT\log^3 D)$  of Algorithm \ref{alg-mi2}.

\begin{figure}
\begin{minipage}[t]{0.48\linewidth}
\centering
\includegraphics[scale=0.21]{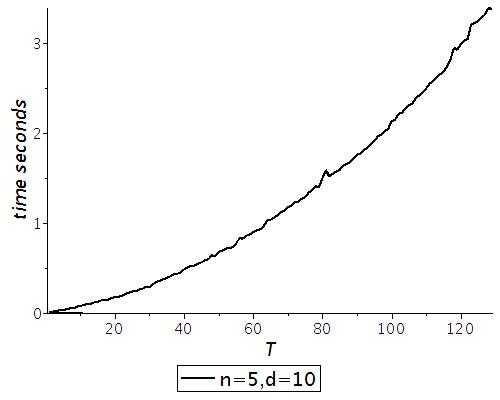}
\caption{Horizontal-axis: $T$ }\label{fig13}
\end{minipage}
\begin{minipage}[t]{0.48\linewidth}
\centering
\includegraphics[scale=0.21]{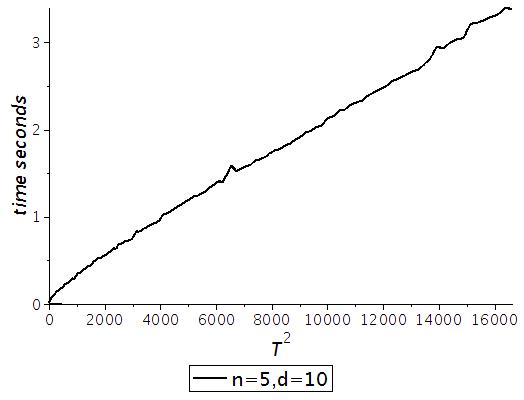}
\caption{Horizontal-axis: $T^2$ }\label{fig14}
\end{minipage}
\end{figure}

\begin{figure}
\begin{minipage}[t]{0.48\linewidth}
\centering
\includegraphics[scale=0.21]{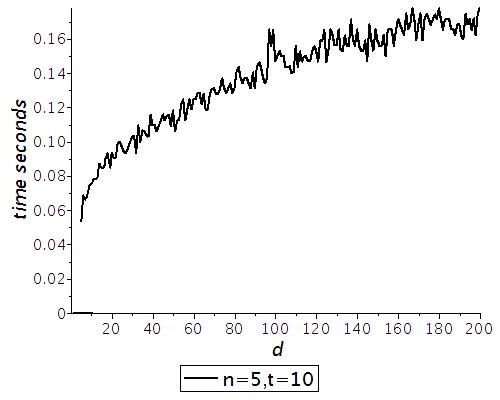}
\caption{Horizontal-axis: $d$ }\label{fig15}
\end{minipage}
\begin{minipage}[t]{0.48\linewidth}
\centering
\includegraphics[scale=0.21]{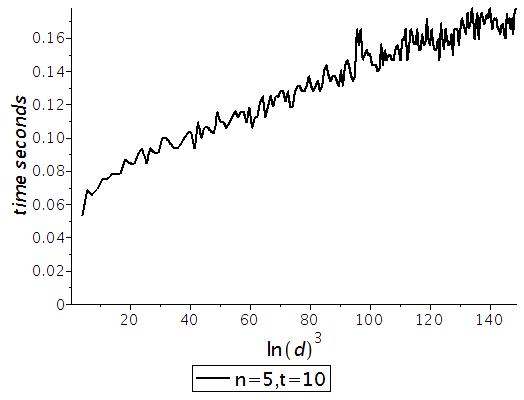}
\caption{Horizontal-axis: $\ln^3 d$ }\label{fig16}
\end{minipage}
\end{figure}

\begin{figure}
\begin{minipage}[t]{0.48\linewidth}
\centering
\includegraphics[scale=0.21]{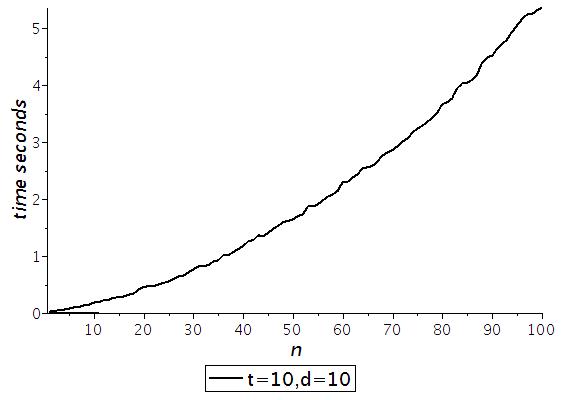}
\caption{Horizontal-axis: $n$ }\label{fig17}
\end{minipage}
\begin{minipage}[t]{0.48\linewidth}
\centering
\includegraphics[scale=0.21]{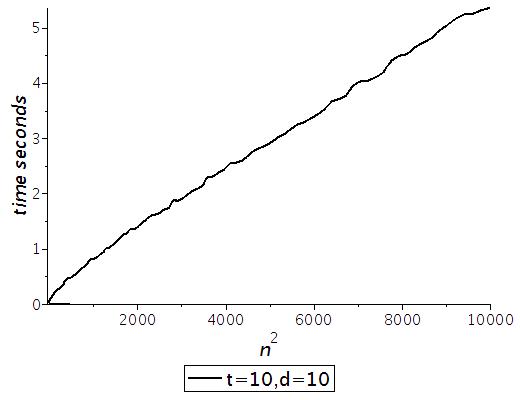}
\caption{Horizontal-axis: $n^2$ }\label{fig18}
\end{minipage}
\end{figure}

For Algorithm \ref{alg-m3}, the relations between the running times and   $T,T^2$ are respectively given in Figures \ref{fig13}, \ref{fig14}, where the parameters $n,d,\mu=1/4$ are fixed. Similarly, the relations between the running times and   $D,\log^3 D,n,n^2$ are respectively given in Figures \ref{fig15}, \ref{fig16}, \ref{fig17}, \ref{fig18}.
These figures show that the practical performance 
is worse than the theoretical complexity bound $O^\thicksim(LnT\log^3 D)$ for Algorithm \ref{alg-m3}, because the logarithm factors omitted in the
soft-Oh analysis have significant impact on the running time.

 \section{Conclusion}
In this paper, we give a new deterministic interpolation algorithm
and two Monte Carlo interpolation algorithms for SLP sparse multivariate polynomials.
Our deterministic algorithm has better complexities than existing deterministic interpolation algorithms in most cases.
Our Monte Carlo interpolation algorithms
are the first algorithms whose complexities are linear in $nT$ and polynomial in $\log D$.
The algorithms are based on several new ideas.
In order to have a deterministic algorithm, we
give a criterion for checking whether a term belongs to a polynomial.
We also give a deterministic method to find a ``good" prime
$p$ in the sense that at least half of the terms in $f$ are not collisions in $f^{\modp}_{(D,p)}$.
Finally, a new Kronecker type substitution is given to reduce multivariate polynomial interpolations to univariate polynomial interpolations.

\end{document}